\documentclass[%
 reprint,
 amsmath,amssymb,
 aps,
]{revtex4-2}

\usepackage{graphicx}% Include figure files
\usepackage{dcolumn}% Align table columns on decimal point
\usepackage{caption}
\usepackage{subcaption}
\usepackage{tikz}
\usepackage{bm}% bold math
\usepackage{float}
\begin{document}

\preprint{APS/123-QED}

\title{Effect of external magnetic field and dust grains on the properties of Ion Acoustic Waves}%

\author{K. Deka}
 %\altaffiliation{}%Lines break automatically or can be forced with \\
\author{R. Paul}%
\author{G. Sharma}
\author{N. Das}
\author{R. Moulick}
\author{S. S. Kausik}
 \email{kausikss@rediffmail.com}
 \author{B. K. Saikia}
\affiliation{Centre of Plasma Physics, Institute for Plasma Research, Sonapur 782402, Assam, India}%

\author{S. Adhikari}
 
\affiliation{
 Department of Physics, University of Oslo, PO Box 1048 Blindern, NO-0316 Oslo,
Norway
}%

\author{O.H. Chin , and C.S. Wong}

\affiliation{Department of Physics, Faculty of Science, University of Malaya, 50603 Kuala Lumpur, Malaysia
}
\date{\today}

\begin{abstract}
An experimental study to investigate the effect of an external magnetic field on the propagation of ion-acoustic waves (IAWs) has been carried out in hydrogen plasma containing two-temperature electrons and dust grains. A low-pressure hot cathode discharge method is opted for plasma production. The desired two electron groups with distinct temperatures are achieved by inserting two magnetic cages with a cusp-shaped magnetic field of different surface field strengths in the same chamber. The dust grains are dropped into the plasma with the help of a dust dropper, which gain negative charges by interacting with the plasma. The IAWs are excited
with the help of a mesh-grid inserted into the plasma. A planar Langmuir probe is used as a detector to detect the
IAWs. The time of flight technique has been applied to measure the phase velocity of the IAWs. The results suggest that in the presence of a magnetic field, the phase velocity of IAWs increases, whereas introducing the dust particles leads to the lower phase velocity. The magnetic field is believed to have a significant effect on the wave damping. This study will aid in utilising IAWs as a diagnostic tool to estimate plasma parameters in the presence of an external magnetic field. Moreover, the study might be useful for estimating the relative ion concentrations in a two positive ion species plasma, as well as the relative concentration of the negative ions in the presence of an external magnetic field.

\end{abstract}

\maketitle

%\tableofcontents

\section{\label{sec:level1}Introduction}

The IAWs are modes of ion oscillations often observed in plasmas. These are low-frequency longitudinal oscillations, which take place with the electrons in the background. These background electrons provide the restoring force essential for the ion oscillations {\cite{hosseini2011}}.
The first theoretical prediction of the IAWs was reported by Tonks and Langmuir {\cite{tonks1929}}. Such waves were experimentally observed by Revans {\cite{Revans1933}} for the first time in a gas discharge tube. The study of IAWs is a very classical area of research in plasma physics, which includes both linear and non-linear modes of the waves {\cite{mahmood2003}}. J J Thomson in 1933 derived the equation of IAWs using the fluid theory {\cite{thomson}}. Damping of the IAWs is one of its intrinsic properties,  classified in two categories viz. Landau damping and collisional damping. Out of these, the Landau damping of the IAWs occurs when the electron temperature ($T_e$) is approximately equal to the ion temperature ($T_i$). The damping and propagation of the IAWs can easily be studied, if the wave in the plasma are excited using an external source.  Wong \textit{et al.} {\cite{wong1964}} excited IAWs with the help of a negatively biased grid and consequently studied Landau damping of the waves. In an investigation on the damping and propagation of IAWs in a plasma containing negative ions, it was observed that when the fraction of negative ions is large, the IAWs split up into two modes of propagation, in which, the faster one undergoes weak Landau damping {\cite{Angelo1966}}. A theoretical and experimental study of the dispersion relation of IAWs in a rare-gas plasma has revealed the existence of two different propagating disturbances {\cite{joyce1969}}. A study of IAWs in a collisionless plasma has shown that the pulse velocity is identical to the phase velocity of the wave {\cite{alexeff1965}}. Tanaca \textit{et al.} have investigated the dispersion of IAWs in a mercury-vapor discharges {\cite{tanaca1966}}, which shows that the cut-off towards the shorter wavelength is $2\pi$ times the Debye length(${\lambda}_D)$.  A non-linear study of IAWs in a three components electron, positron, and ion plasma investigated the formation of ion-acoustic solitons {\cite{popel1995}}. It was found that the presence of positrons in a multicomponent plasma reduces the amplitudes of the solitons. The shock formation of non-linear IAWs in a collisionless unmagnetized plasma has shown that the wave mach number increases with the excitation amplitude {\cite{taylor1970}}. The presence of two-temperature electrons in plasma may have a significant effect on its various applications {\cite{sharma2020}}. IAWs are also studied in a plasma containing two-temperature electrons. The non-linear propagation of IAWs in a collision-dominated two-electron temperature plasma has been explored by Shukla \textit{et al.} {\cite{shukla1976}}. The results of the study suggest that in such an environment, the IAWs develop shock-like structures. In a two-electron temperature plasma, the study of IAWs reveal that the group with lower energy has the dominant effect on the propagation of IAWs {\cite{Jones1975}}.

 In the presence of dust particles, the IAWs get modified as Dust Ion Acoustic Waves (DIAWs). The DIAWs propagate in the range of frequency $kv_{th,i}<<\omega<<kv_{th,e}$, where $\omega$, $k$, $v_{th,i}$ and $v_{th,e}$ are the wave angular frequency, wave number, ion thermal velocity and electron thermal velocity respectively. The DIAWs travel with higher phase velocity than the usual IAWs {\cite{merlino1997}}. Barkan \textit{et al.} {\cite{barkan1996}} studied DIAWs in a Q-machine. It was observed that in the presence of dust grains, the phase velocity of IAWs increases and Landau damping decreases. Nakamura \textit{et al.} {\cite{nakamuraPRL}} studied linear and non-linear DIAWs in a homogeneous unmagnetized dusty plasma. The results of the study are significant in the sense that for the linear regime of the wave, the phase velocity and damping increase with the increase in dust density. Dispersion relation of DIAWs was derived along with the rate of Landau damping by a kinetic approach {\cite{hadi2019}}. The non-linear mode of propagation of the DIAWs has been explored by many researchers {\cite{nakamura2001,maitra2006}}. One of such studies on the formation of ion-acoustic shock waves in a dusty plasma has thrown some light on the fact that in the absence of the dust grains, the ion oscillations lead to a train of solitons in the plasma {\cite{nakamura2002}}. In another study the non-linear DIAWs in a four-component plasma were investigated  {\cite{farooq2017}}. The results show that the amplitude and width of the solitary waves are affected by the presence of superthermal electrons. 
 
 Although there is a significant amount of literature dealing with the various aspects of IAWs in plasma, the field lacks the proper understanding of linear IAWs in presence of an external magnetic field. The present work deals with the effect of an external magnetic field on the IAWs. The IAWs are widely used as a diagnostic tool to determine the electron temperature and ion mass of plasma. In addition, they are used to estimate the relative ion concentrations in a two positive ion species plasma, as well as the relative concentration of the negative ions present in the plasma {\cite{hershkow2008,kakati2017, Saikia2014}}. Negative ions are the prime participants for the Neutral Beam Injection (NBI) in ITER tokamak. The production of the negative ions in plasma involves two methods, volume and surface production. The advantage of the volume process lies in the fact that the negative ions are produced in an isotropic manner. On the other hand, the surface production has better efficiency over the volume process {\cite{kakati2017}}. It has been shown that the densities of negative ions increase in the presence of external magnetic fields {\cite{santoso2015}}. In a low pressure plasma, a novel surface-assisted volume negative hydrogen ion source has been developed by using cesium (Cs) coated Tungsten (W) dust {\cite{kakati2017}}.  In these studies, the addition of an external magnetic field may lead to an efficient negative ion source by enhancing the density of the negative ions. In such works, the IAWs in the presence of an external magnetic field may be utilized as a diagnostic tool to estimate the density of the negative ions. The present study has been carried out in a hydrogen plasma so that IAWs can be used as a reliable diagnostic tool for confirming the presence of negative ions and consequently estimating their density.
 
 \begin{figure}
\includegraphics[width=8.6cm]{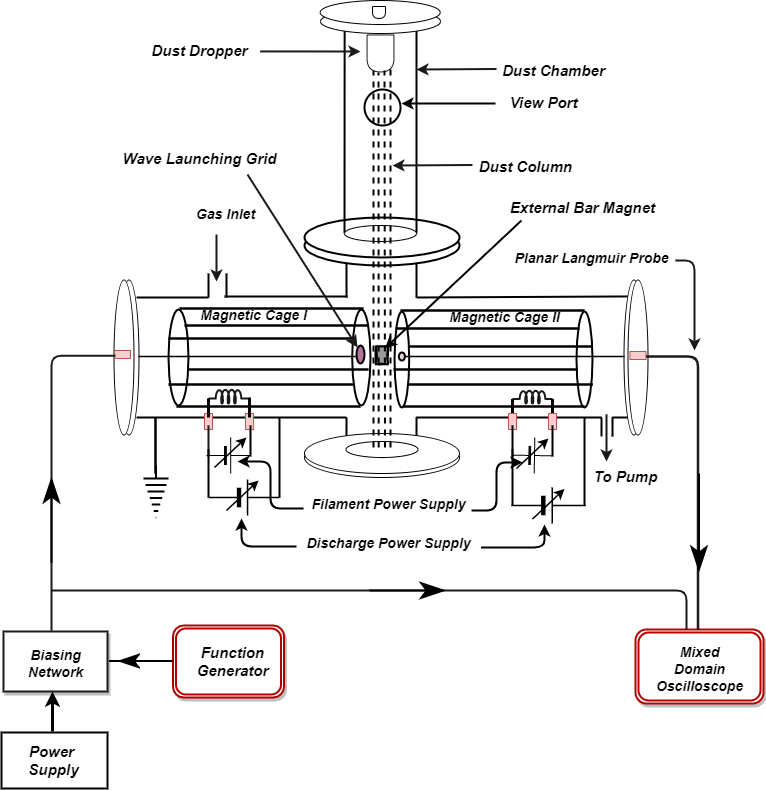}
\centering
\caption{Schematic diagram of the experimental set up.}
\end{figure}
 
 The motivation behind the present work is to understand how the presence of an external magnetic field affects the propagation and damping properties of linear IAWs.
Additionally, the propagation of IAWs depends upon the temperature of the electrons in the plasma. It is a well-known fact that in a plasma with two distinct groups of electrons, the colder electrons have a dominant effect on the IAWs. The effect is prominent even if the plasma contains only a small fraction of cold electrons. The optical techniques often fail to detect the presence of such cold electrons in the plasma. For those plasmas containing hot electrons as the dominant group, the IAWs acts as a suitable diagnostic tool to determine the presence of cold electrons {\cite{Jones1975}}. Moreover, dust charging study in a two-electron temperature plasma has revealed secondary electron emissions (SEE) from Tungsten (W) dust grains {\cite{paul2021}}. The study shows that the SEE from the dust grains results in an increase in the density of cold electrons, which is expected to have a significant effect on the propagation of IAWs. The present study aims to investigate the role of the two electron groups on the characteristics of IAWs in the presence of dust and an external magnetic field. 

The paper has been categorized as follows. In section 2, the production mechanism of the plasma, IAWs excitation and detection have been described. Analysis of the results observed in the experiment is presented and discussed in section 3. The conclusions of the study are drawn in section 4.

\section{Experimental  set up}
 The experimental study of the IAWs has been carried out in a cylindrical stainless steel chamber consisting of horizontal and vertical chambers.
 The experimental chamber looks like an inverted T, as shown in Fig. 1. The plasma is produced in the horizontal chamber, having a length of 100 cm and a diameter of 30 cm. On the other hand, the vertical chamber is used as the dust dropping unit with a height of 72 cm and a diameter of 15 cm. The chamber is evacuated to a base pressure of $1.5\times10^{-5}~mbar$ using a diffusion pump backed by a rotary pump. The working pressure of the hydrogen gas is maintained at $1.7\times10^{-4}~mbar$. Hot cathode filament discharge method has been used to produce the plasma. The primary electrons emitted by the filaments are  accelerated by a biasing voltage in the range of 80-90 Volts, which participate in the ionization of the gas and produce secondary electrons. These secondary electrons are accelerated by the biasing voltage and produce further ionization. Thus a sustainable discharge condition is achieved. Plasma with two distinct groups of electrons having different temperatures is achieved by inserting two magnetic cages of different surface field strengths (Magnetic Cage I: 1.2 kG made of Strontium Ferrite, and Magnetic Cage II: 3.5 kG made of a Samarium Cobalt) inside the chamber, as depicted in Fig. 1. Different discharge currents are maintained in the two cages by suitably controlling the filament currents. The discharge current in the Cage I is fixed at $0.5~A$, whereas in the Cage II it was varied from $1~A-3~A$ at a step of $0.5~A$. Thus, two different plasmas are produced in the individual cages. The plasmas so produced diffuses at the junction of the two cages, and consequently the plasma at the junction consists of two-electron groups with different temperatures {\cite{sharma2021}}. The semi-logarithmic plot of the \textit{I-V} having two distinct slopes, as displayed in Fig. 2, confirms the presence of two different electron species with distinguishable temperatures.

\begin{figure}
\includegraphics[width=8.6cm]{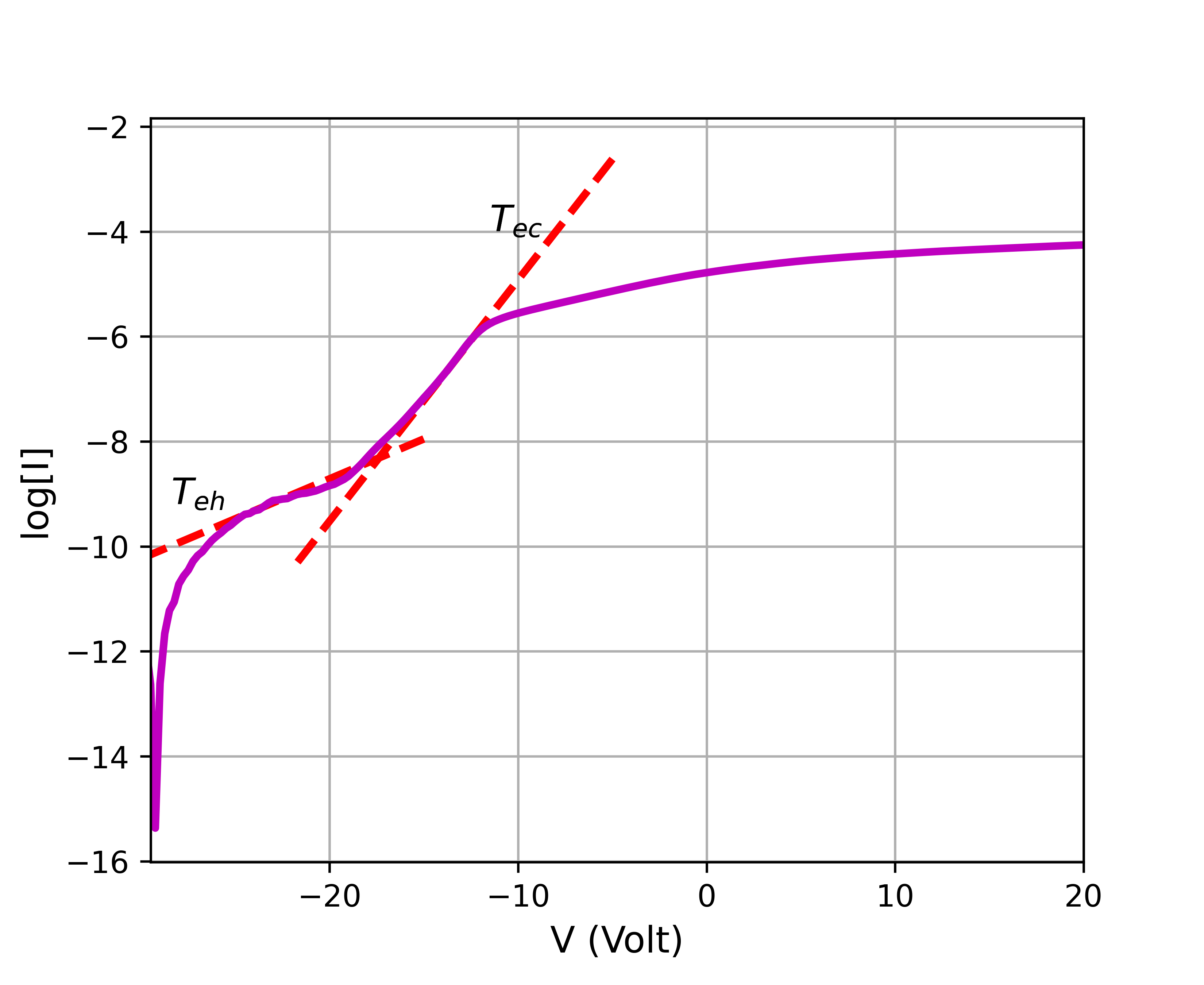}
\centering
\caption{\textit{I-V} characteristics in semi-log scale depicting the presence of two electron temperatures}
\end{figure}

 The upper and lower slopes give the electron temperatures of cold ($T_{ec}$) and hot ($T_{eh}$) components respectively {\cite{pustylnik2006, pilling2007}}.

The effective electron temperature of the plasma is given by {\cite{Jones1975}}
 \begin{equation}
 T_{eff}={n_eT_{ec}T_{eh}}/({n_{eh}T_{ec}+n_{ec}T_{eh}})
 \end{equation}
 where, $n_e$, $n_{ec}$, and $n_{eh}$ are the total plasma density, density of cold, and hot components of the electrons respectively.
 
 The phase velocity of the IAWs in a two-electron temperature plasma {\cite{Jones1975}} takes the form
 
 \begin{equation}
 \omega/k = \sqrt{k_BT_{eff}/m_i}
 \end{equation}
where, $\omega$, $k$, $k_B$, $T_{eff}$, and $m_i$ are angular frequency, wave number, Boltzmann constant, effective electron temperature, and mass of the ion respectively.

In this experiment, the IAWs have been excited by inserting a mesh-grid into the plasma column {\cite{wong1964,barkan1996}}, which is made of stainless steel, having a diameter of $5~cm$ with transparency of $\approx 75\%$.  For appreciable density modulation, the inner-wire spacing of the grid should be comparable to the Debye length and ion gyro-radius {\cite{barkan1996,mamun2001}}. In the present context, the Debye length and ion gyro-radius are $\approx0.1~mm$ and $0.115~mm$ respectively. Therefore, the inner-wire spacing of the grid is $\approx0.5~mm$, and its surface is perpendicular to the axis of the plasma column. For launching the IAWs, a transistor-based biasing circuit was designed and fabricated, and a sinusoidal voltage, having a frequency in the range of $80-120~kHz$ with peak-to-peak voltage 4-5 V was applied to it.
The applied signal creates a density perturbation, and consequently, it propagates through the plasma as IAWs. A planar Langmuir probe (floating) having a diameter of approximately $9~mm$ is used to detect the waves. The grid and probe are maintained at a distance of $2~cm$ from each other. The phase shift and amplitude of the detected waves are measured and recorded with the help of an Oscilloscope (MDO3054, $500~MHz$). A radially movable bar magnet with a surface field strength of 3.5 kG is introduced into the plasma column in such a way that the magnetic field is perpendicular to the axis of the wave launching grid and the receiver.  Fig. 3 (a) describes the magnetic field profile of the external magnet with its position from the grid-probe axis. It can be seen that the magnetic field strength decreases as the magnet is moved away from the axis of the grid and probe.

\begin{figure}
\begin{subfigure}[b]{0.44\textwidth}
         \centering
         \includegraphics[width=\textwidth]{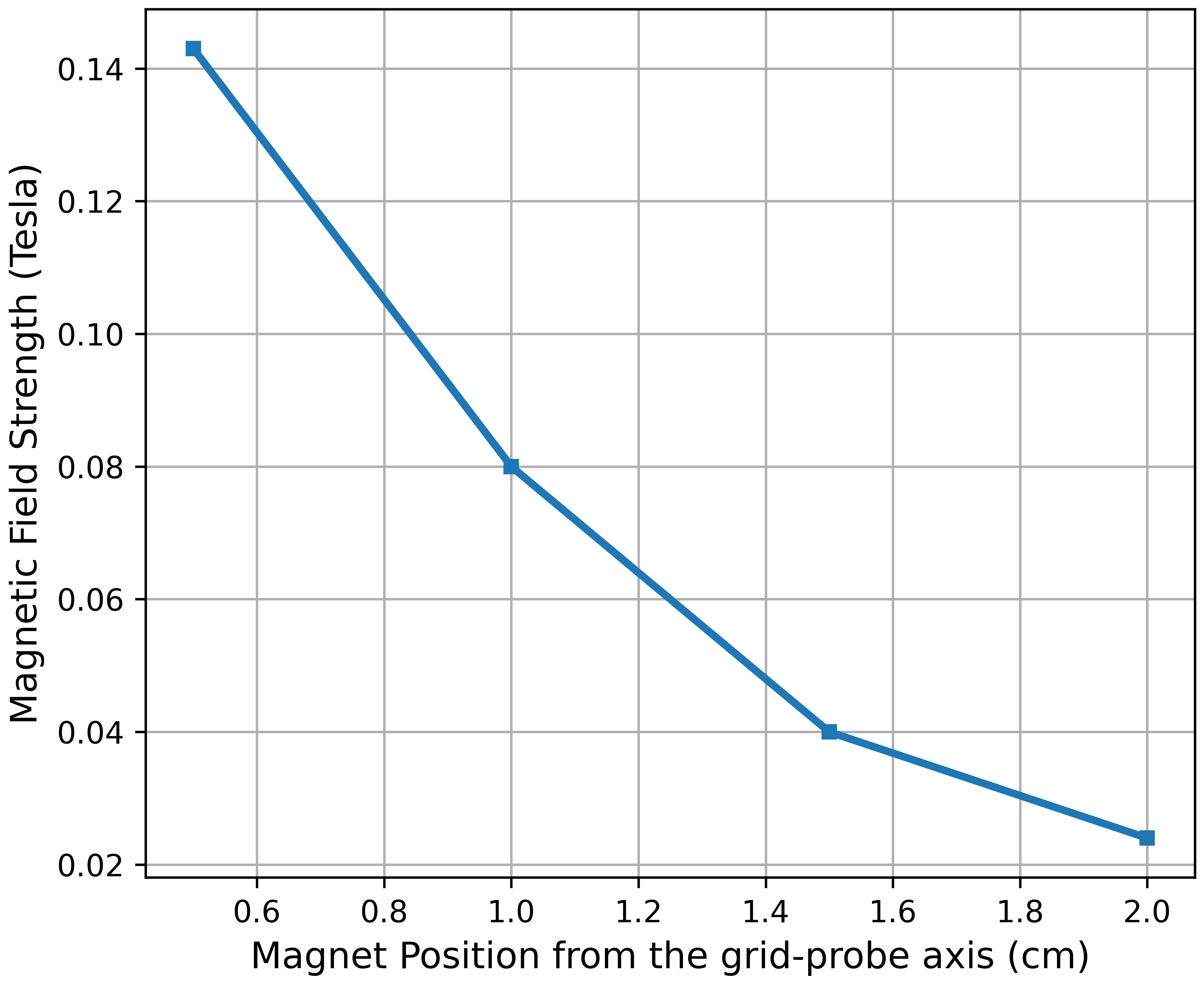}
         \caption{}
\end{subfigure}
\hfill
\begin{subfigure}[b]{0.44\textwidth}
         \centering
         \includegraphics[width=\textwidth]{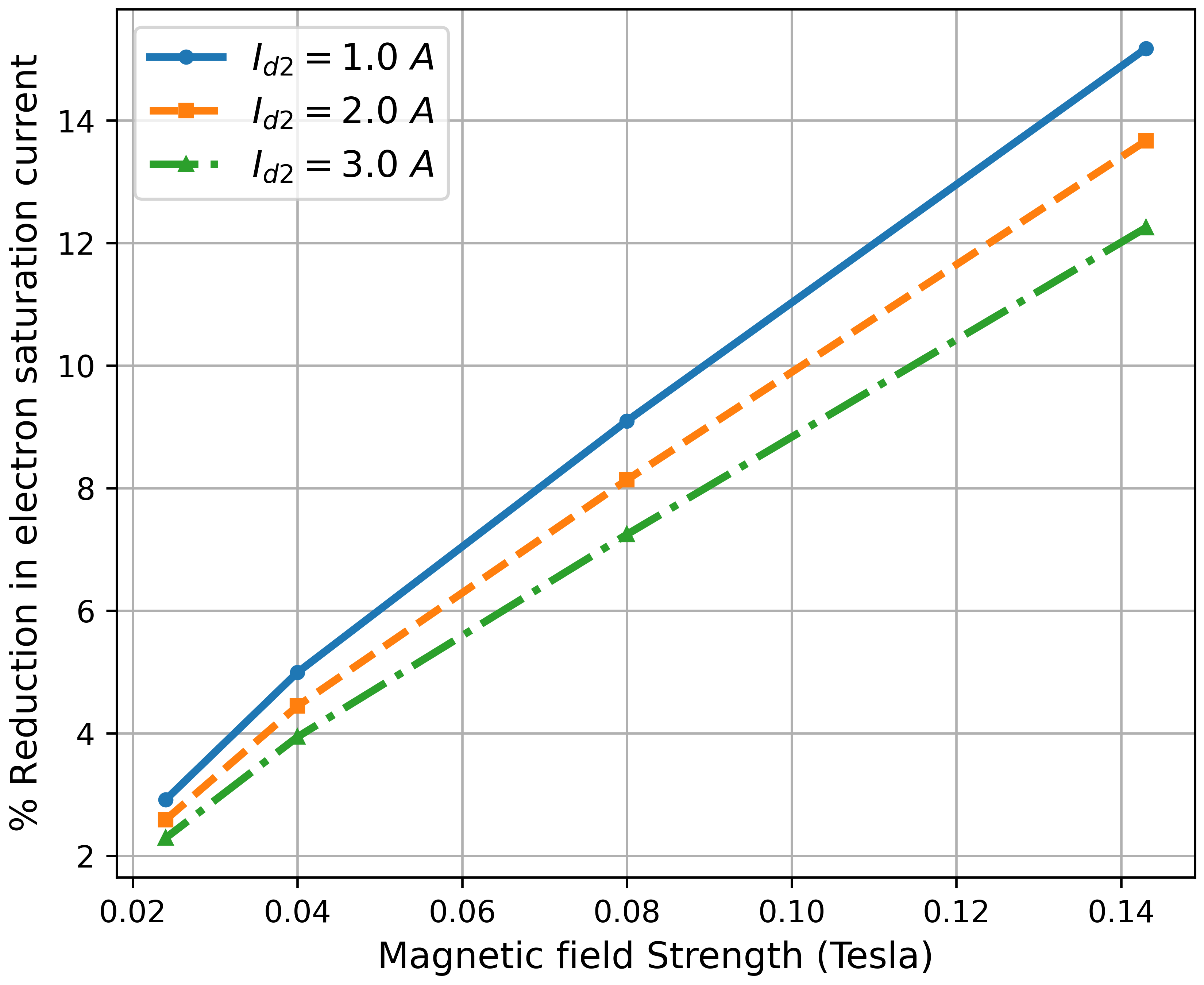}
         \caption{}
\end{subfigure}
\caption{(a) Magnetic field profile of the external magnet with its position from the grid-probe axis. (b) Percentage of reduction in electron saturation current to the cylindrical Langmuir probe in the presence of the external magnetic field for a constant discharge current of $I_{d1}=0.5~A$ in Cage I, and a varying discharge currents $I_{d2}$ in Cage II.}
\end{figure}

Tungsten (W) dust grains are introduced into the plasma with the help of a dust dropper located at the top of the vertical chamber of the experimental setup as depicted in Fig. 1. The dust dropper consists of a cylindrical vessel with a mesh at the bottom surface through which the dust particles fall into the plasma. A d.c. motor is attached to the small dust-containing cylindrical chamber in such a way that whenever the motor rotates, the dust vessel undergoes vibration, and consequently, the dust particles fall through the mesh and enter into the plasma column. These dust grains, in general, acquire a negative charge by interacting with the plasma.

\subsection{Estimation of Electron Saturation Current using Langmuir probe in the Presence of External Magnetic Field:}

The electron current collected by the Langmuir probe is suppressed in the presence of an external magnetic field. The reduction in electron current to the probe is described by the reduction factor \textit{r} {\cite{pitts1990, stangeby1982, tagle1987}}
\begin{equation}
r= 2{\lambda}_{mfp}\frac{\sqrt{\alpha}(1+\tau)}{d}
\end{equation}
where,
\begin{equation*}
\alpha={D_{\perp}}/{D_{\parallel}}~~~~ \tau=T_i/T_e.
\end{equation*}

${\lambda}_{mfp}$ $\rightarrow$ is the momentum loss mean free path of the electrons to the ions.

~~~~$D_{\perp}$ $\rightarrow$ Diffusion co-efficient perpendicular to magnetic field.

~~~~$D_{\parallel}$ $\rightarrow$ Diffusion co-efficient parallel to magnetic field.

~~~~~${d}$ $\rightarrow$ radius of the probe.

The reduced value of electron saturation current is given by {\cite{tagle1987}}
\begin{equation}
I^e_{sat}=\frac{1}{4}n_0e\Bar{c_e}A\bigg(\frac{r}{1+r}\bigg)
\end{equation}

\begin{equation*}
\Bar{c_e}=\sqrt{\frac{8kT_e}{{\pi}m}}
\end{equation*}

where, $n_0$, $\Bar{c_e}$, and $A$ are the undisturbed plasma density, random electron thermal velocity, and probe area respectively. However, the ion and electron currents to a floating probe are insensitive to the current reduction factor $r$, and size of the probe $d$ in the presence of a magnetic field {\cite{stangeby1982, stanojevic1994}}. In this work, a planar Langmuir probe at floating potential has been employed for the detection of the IAWs. So, there will be no concern on the validity of the orbital-motion-limited (OML) theory for the wave detection in the presence of the magnetic field by the planar Langmuir probe at floating potential.

A cylindrical Langmuir probe having a length of $3~mm$ and diameter of $0.3~mm$ is used to determine the plasma parameters.
The percentage of reduction in electron saturation current, for the present context, in an external magnetic field, for different values of the discharge currents is represented in Fig. 3(b). It is observed that the maximum and minimum value of reduction is approximately $15\%$, and $2.2\%$ respectively. The figure also depicts that with the increase of magnetic field strength, the percentage of reduction in electron saturation current increases, but it decreases with the increase in discharge current.

An attempt has been made to compensate for the reduction in electron saturation current to the Langmuir probe in the presence of the magnetic field by adding a correction factor, equivalent to the amount of reduction to the estimated value of current. The corresponding saturation current has been used to determine the value of the electron density.

\subsection{Estimation of Electron temperature using Langmuir probe in the presence of Magnetic field:}

Erents \textit{et al.} {\cite{erents1986}} and Tagle \textit{et al.} {\cite{tagle1987}} estimated the value of electron temperature on the plasma boundary of a JET tokamak by applying a non-linear fit of the form
\begin{equation}
I = I^i_{sat}\bigg[1-\exp\bigg\{\frac{e(V-V_f)}{kT_e}\bigg\}\bigg]
\end{equation}
to the region below floating potential $(V<V_f)$, as they found a deviation of the \textit{I-V} from the exponential behavior above the floating potential. In such a scenario, the \textit{I-V} above the floating potential falsely depicts high electron temperature. In the present work, a similar method as that employed by Erents \textit{et al.} and Tagle \textit{et al.} was tried for the estimation of the electron temperatures, although the \textit{I-V} in the presence of the magnetic field does not show any discrepancy from the exponential behavior. However, the method has been found inappropriate for the present context, as it yields electron temperatures that are even much higher than the hot component of the plasma under consideration. Therefore, a different method has been adopted to determine the electron temperature.  The first derivative of the \textit{I-V} is used as the standard method for the evaluation of plasma potential. But, in the presence of a magnetic field, the peak obtained from the first derivative of the I-V is indistinct, leading to some ambiguity in the determination of plasma potential, which is more prominent for a high magnetic field.
Hence, this method is also inadequate in the presence of a magnetic field. For such cases, the intersection method is more appropriate for the estimation of plasma potential. The value of plasma potential $(V_p)$, from the semi-logarithmic plot of the \textit{I-V}, is estimated from the intersection of the lines obtained out of {\cite{maria2018, merlino2007}}

\begin{enumerate}
 \item extrapolation  of the region between floating potential $(V_f)$ and plasma potential $(V_p)$.
 \item extrapolation of the saturation region of the \textit{I-V}.
\end{enumerate}

Now, after the correct estimation of the plasma potential, the slope of the linear region between plasma potential (obtained from the intersection method) and floating potential yields the electron temperature. 

\section{Results and Discussion}
%\subsection{Two electron temperature plasma}

\subsection{Variation of Phase Velocity of IAWs:~~~~~~~~~~~~~~~~~~~~~~~~~~~~~~~~~~~~~~~~~~~~~~~}
Fig. 4 (a) represents the typical waveforms of the applied and the received signals of the IAWs. The phase difference and damping are clearly visible from the figure. 

\begin{figure}
\begin{subfigure}[b]{0.44\textwidth}
         \centering
         \includegraphics[width=\textwidth]{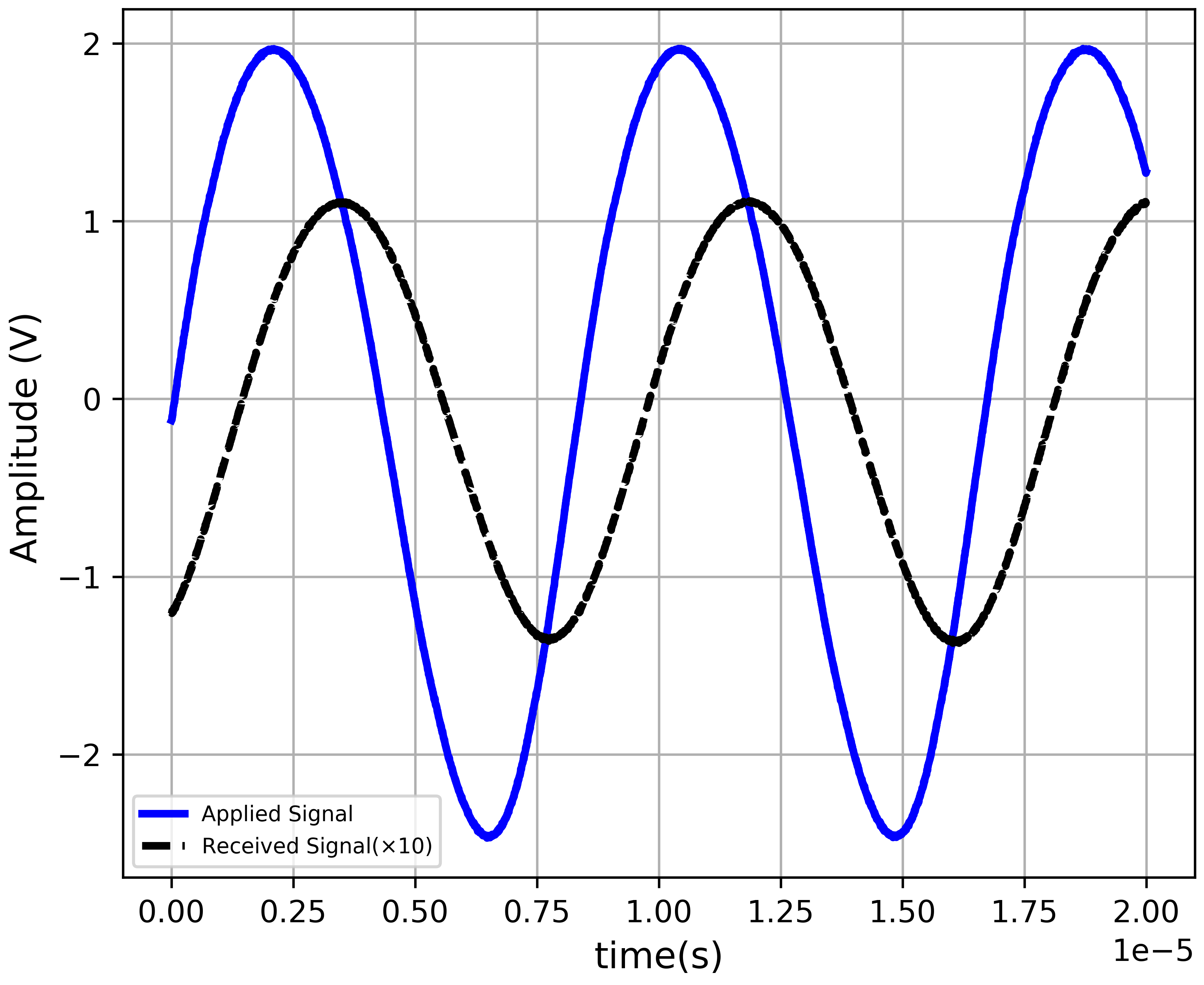}
         \caption{}
\end{subfigure}
\hfill
\begin{subfigure}[b]{0.44\textwidth}
         \centering
         \includegraphics[width=\textwidth]{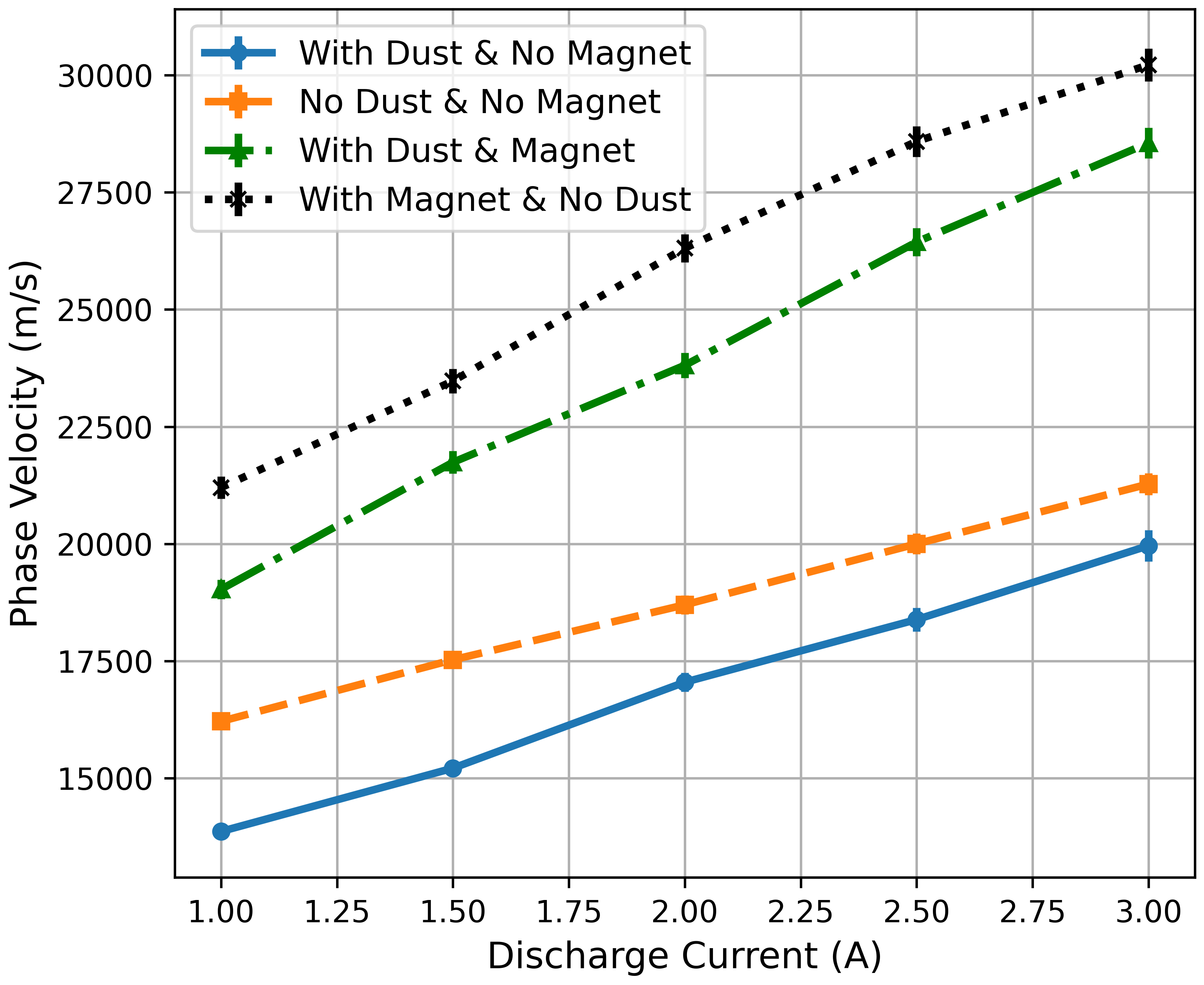}
         \caption{}
\end{subfigure}
\caption{(a) Wave form of applied and received signal of the wave representing the phase difference and damping.  (b) Phase Velocity of IAWs at a constant discharge current of $I_{d1}=0.5~A$ in Cage I and different values of discharge currents $I_{d2}$ in Cage II.}
\end{figure}

Fig. 4 (b) depicts the phase velocity of IAWs in a two-electron temperature plasma. The phase velocity appears to grow as the plasma discharge current increases. The temperature and density of electrons rise with the discharge current as portrayed in Fig. 5. The increase in electron density tends to provide more shielding to the electric field of the ion bunches, but the subsequent increase in electron temperature reduces the shielding effect. As a result, the phase velocity of the wave will increase.

\begin{figure}
\includegraphics[width=8.6cm]{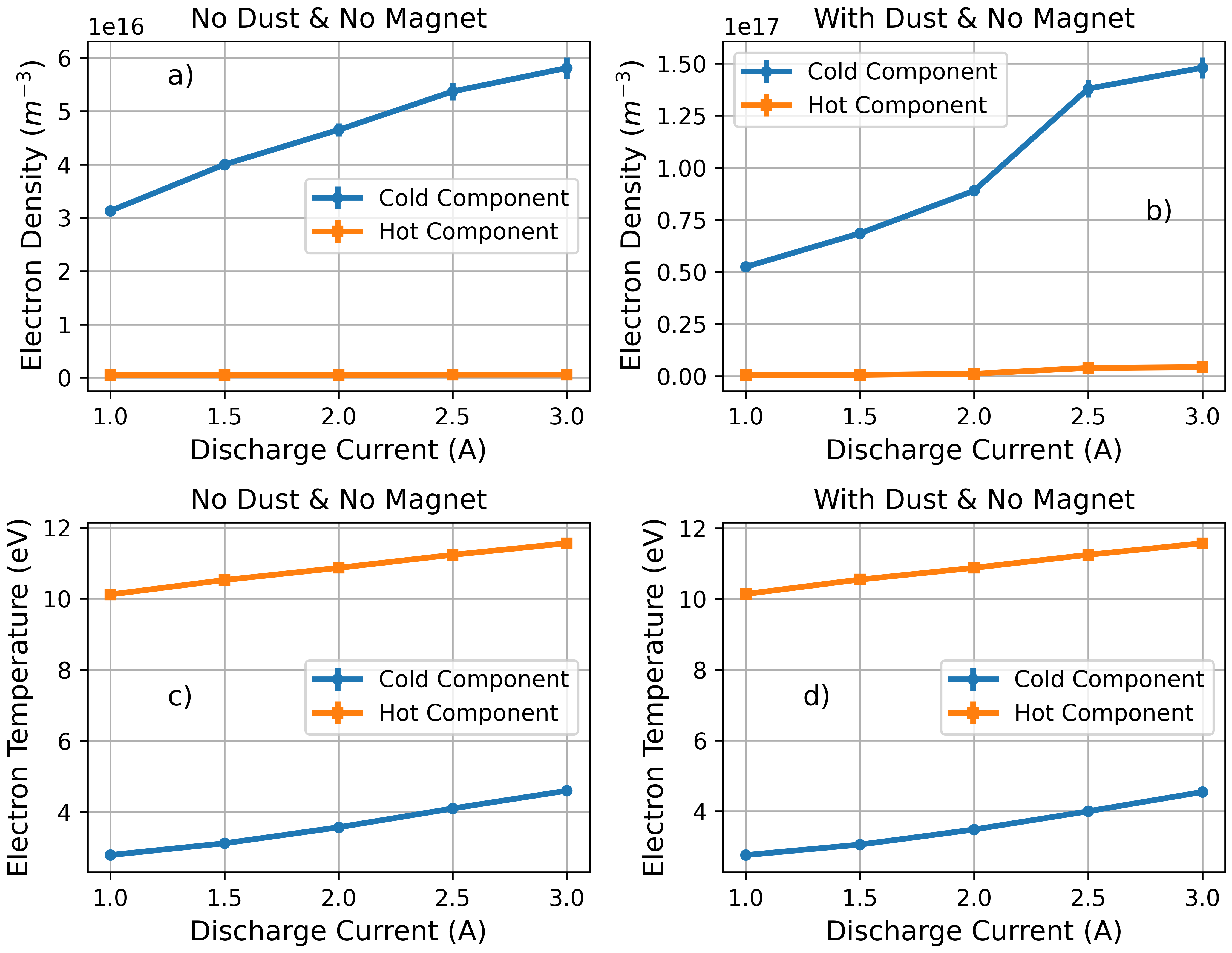}
\centering
\caption{ $(a), (b):$ Plots of Electron densities, and  $ (c),  (d):$ plots of electron temperatures at a constant discharge current of $I_{d1}=0.5~A$ in Cage I, and different values of discharge currents $I_{d2}$ in Cage II  for pure plasma and in the presence of dust particles respectively.}
\end{figure}

In a single electron temperature plasma, the phase velocity of the IAWs increases in the presence of dust particles due to the attachment of plasma electrons on the dust grains, which is visible in Fig. 6 (a). But in this case, the phase velocity of the IAWs decreases in the presence of dust grains, as shown in Fig. 4 (b) and 6 (b). The justification for it is put forward with the help of Fig. 5. It shows that in the presence of dust grains, there is a significant rise in the density of the cold component of electrons in a two-electron temperature plasma due to the secondary electron emissions from the dust, caused by the electron group with higher temperature {\cite{mamun2003, paul2021}}. Besides this, the electron temperature does not change significantly. Due to the rise in density, the shielding of the electric field associated with the bunching of ions increases, and hence the phase velocity of the wave decreases.

\begin{figure}
\includegraphics[width=8.6cm]{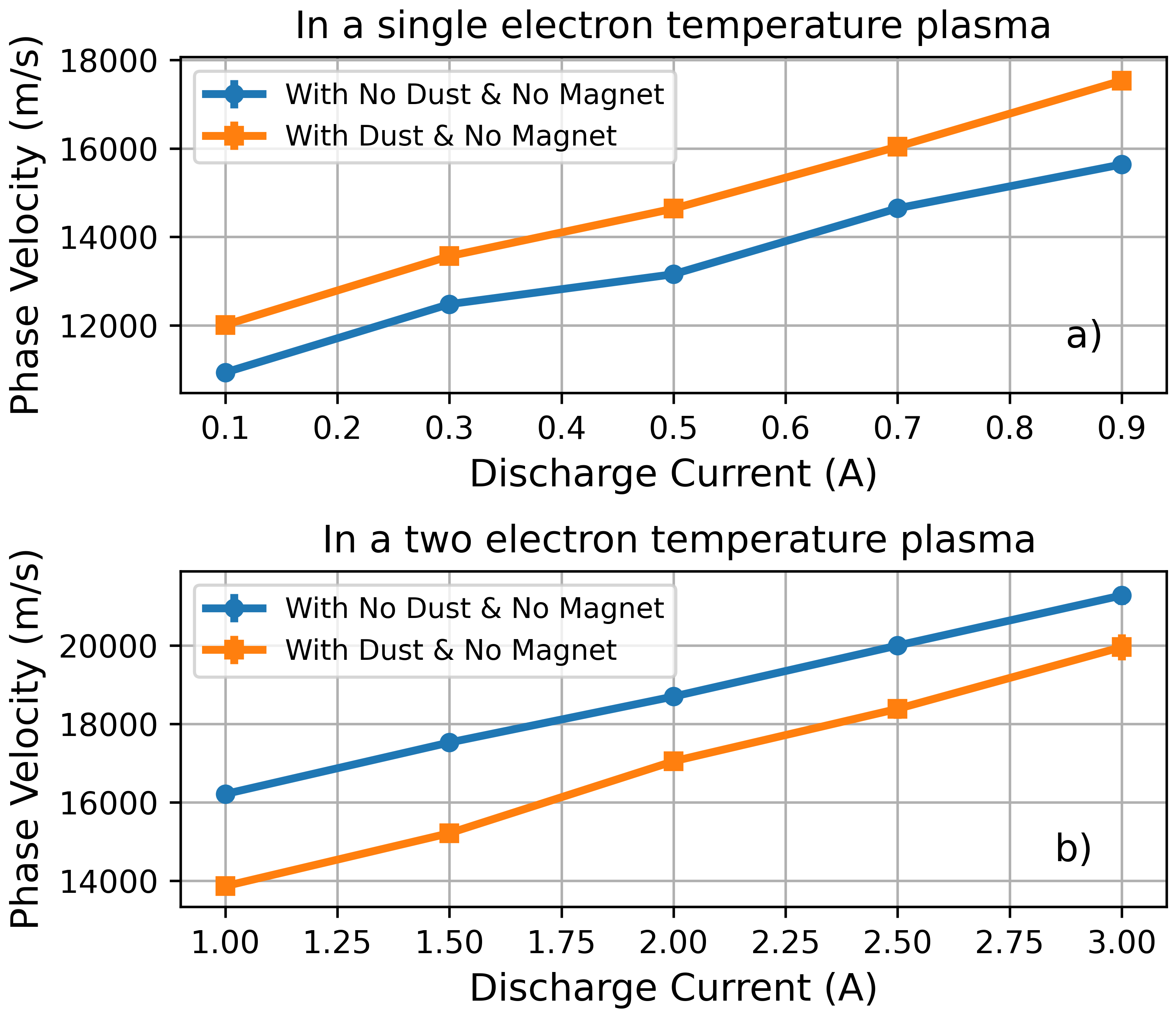}
\centering
\caption{a) Phase Velocity of IAWs in a single electron temperature plasma with discharge currents, and b) phase velocity of IAWs in a two electron temperature plasma for a constant discharge current of $I_{d1}=0.5~A$ in Cage I and different values of discharge currents $I_{d2}$ in Cage II.}
\end{figure}

The secondary electron emissions from the dust grains can also be observed from the \textit{I-V} curves shown in Fig. 7. Under normal circumstances, the electron saturation current to the probe drops in the presence of dust grains, indicating that electrons are attached to the dust grains. The electron saturation current can increase only when there are additional electrons in the plasma. The only possible source of these additional electrons, at fixed plasma conditions, is the secondary emissions from the surface of dust grains. Thus the study of IAWs can be considered as an alternative method for the confirmation of secondary electron emissions from the dust grains in a two-electron temperature plasma that contains high-energetic electrons as we stated in the beginning.

\begin{figure}
\includegraphics[width=8.6cm]{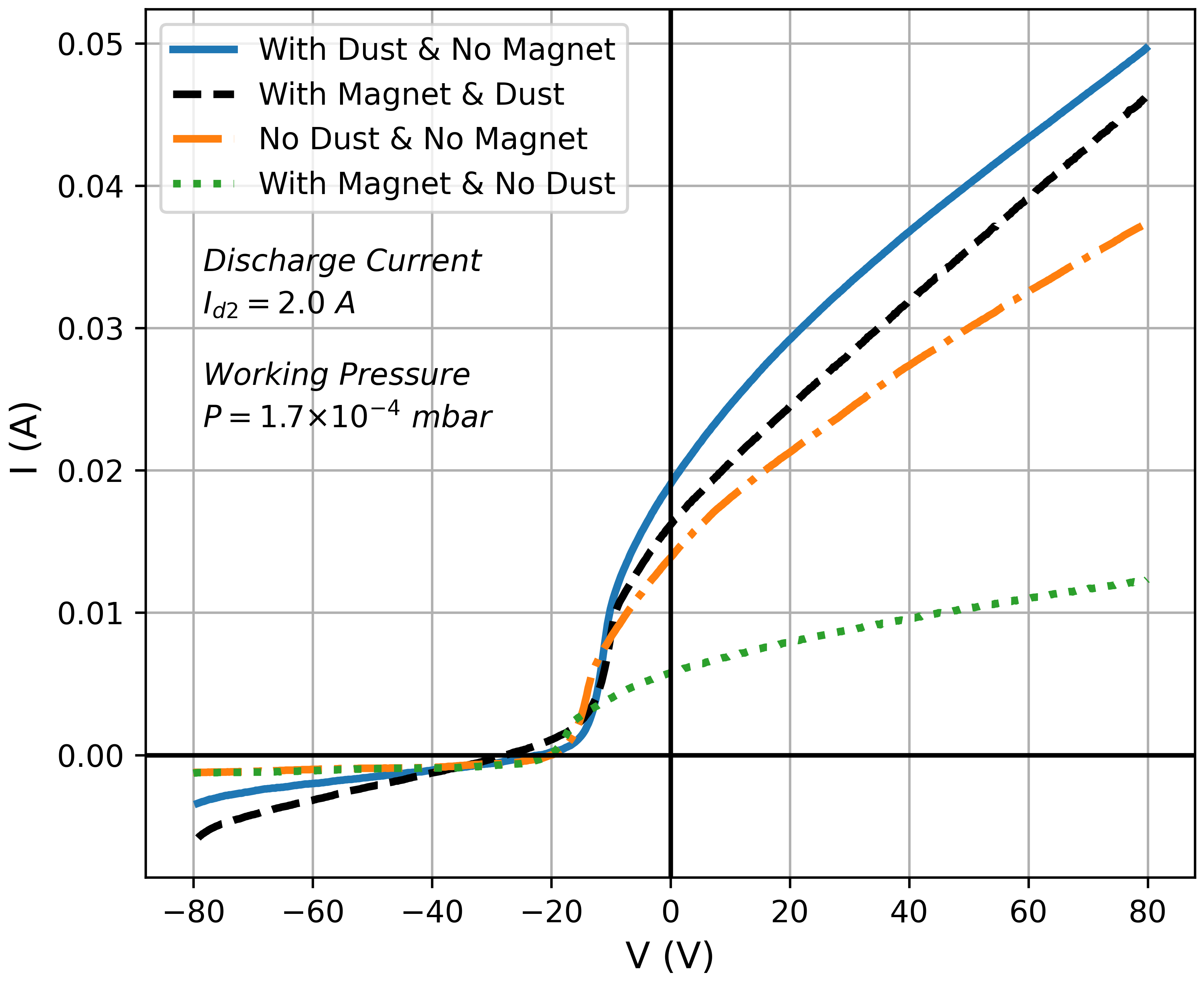}
\centering
\caption{Langmuir probe \textit{I-V} characteristics of plasma at a discharge current of $I_{d1}=0.5~A$ in Cage I, and $I_{d2} = 2.0~A$ in the Cage II.}
\end{figure}

In the presence of an external magnetic field, the electrons participating in the propagation of IAWs may undergo two types of motions depending on the direction of electron velocity,
a) cyclotron motion about the field direction, and b) helical motion along the field. The electrons undergoing the cyclotron motion are confined by the magnetic field and are prevented from freely shielding the electric field created by the bunching of ions. Further, the electrons belonging to the helical trajectory are drifted away from the wave direction, which as a consequence reduces the number of electrons participating in the shielding process. This subsequently leads to the increase in the phase velocity of IAWs.  The effective electron temperature increases only slightly in the presence of the magnetic field, and hence can not play a significant role in the increase in phase velocity. The facts have been depicted in Figs. 8 (a) and (b).

On introducing the dust particles into the plasma, the electrons collide with the dust grains under the effect of the magnetic field. In such collisions, three processes may take place, a) attachment of the electrons on the dust surface, b) scattering of the electrons from the surface of dust, and c) secondary electron emissions from the dust grains. The gyro-radius of the electron in this case is greater than the size of the dust particles, which may enhance the scattering of the electrons from the dust and reduce the secondary electron emission from the dust surface by the high energy electrons. The scattering of the electrons from the dust surface will further lead to two possibilities, a) instantaneous modification of the gyro-orbit of the charged particle, b) restriction of the mobility of electrons along the field direction. The density of electrons in the presence of dust and magnet is higher than in the absence of dust and magnet (\textit{i.e} magnet alone) due to secondary emissions and scattering. There is no significant change in the effective electron temperature in the presence of dust $\&$ magnet as depicted in Fig. 8 (a). Therefore, this increase in electron density enhances the shielding of the electric field of ion bunches and hence, the phase velocity decreases.

\begin{figure}
\begin{subfigure}[b]{0.44\textwidth}
         \centering
         \includegraphics[width=\textwidth]{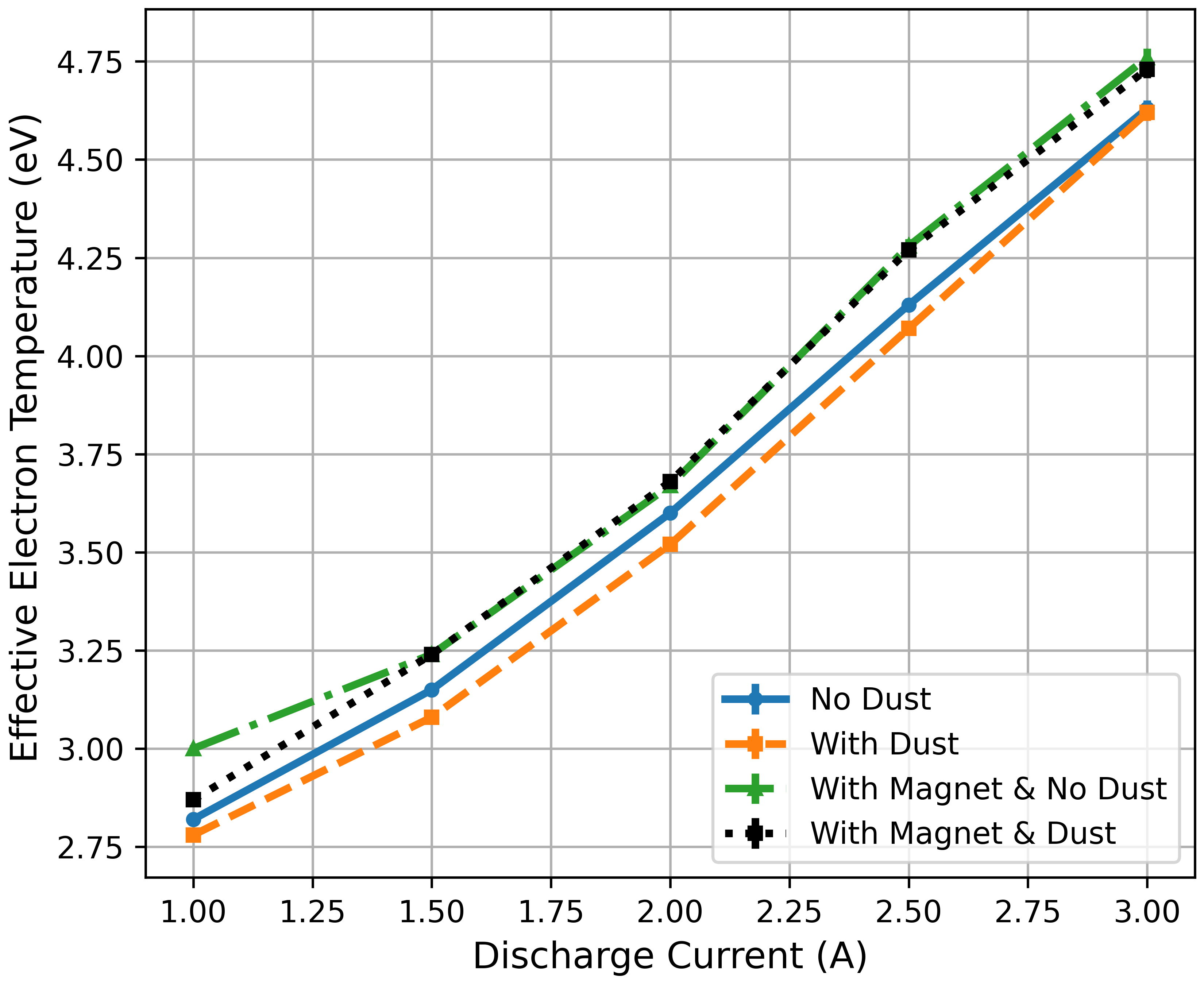}
         \caption{}
\end{subfigure}
\hfill
\begin{subfigure}[b]{0.44\textwidth}
         \centering
         \includegraphics[width=\textwidth]{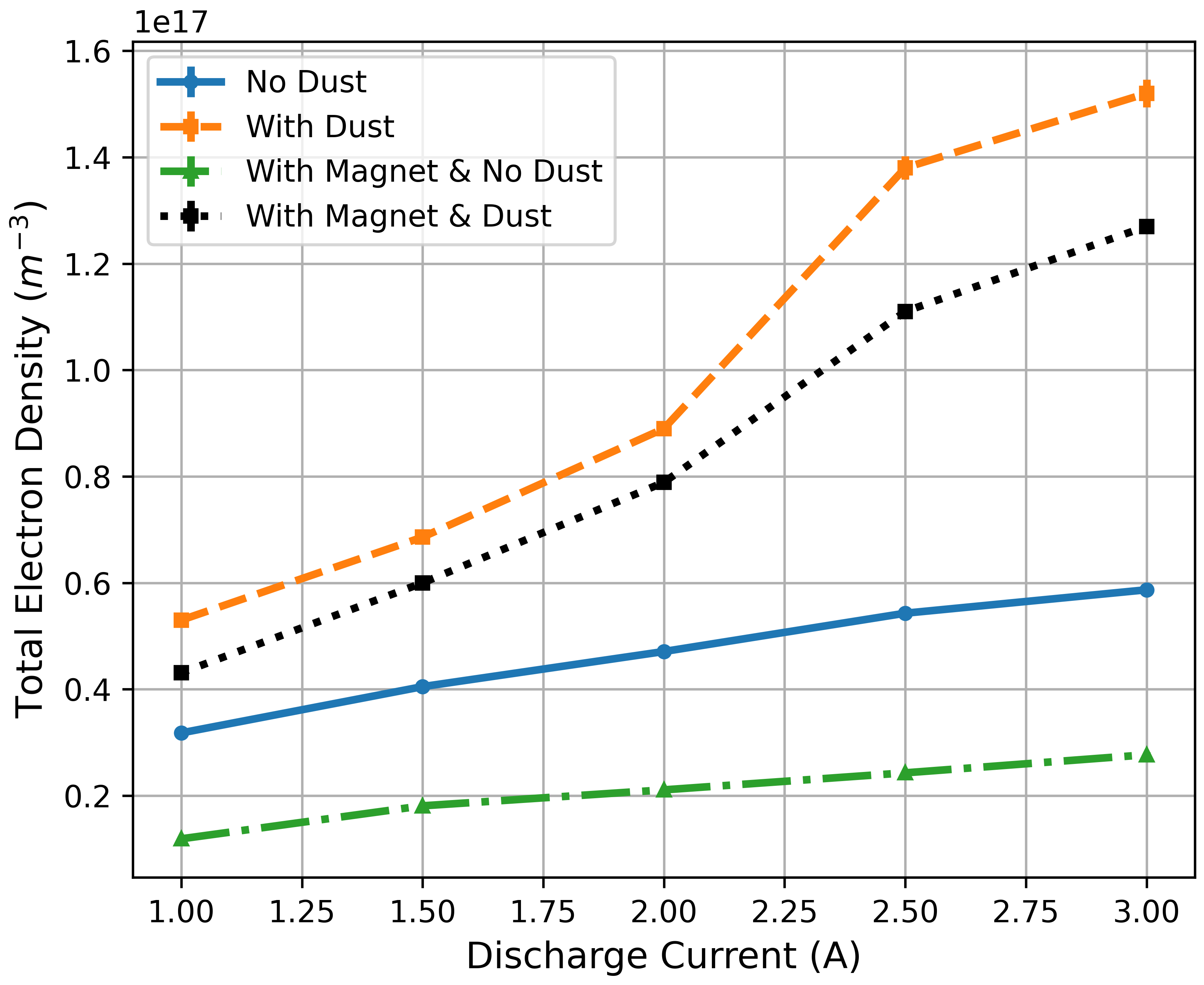}
         \caption{}
\end{subfigure}
\caption{(a) Effective temperature of electrons for constant discharge current of $I_{d1}=0.5~A$ in Cage I and with a variable discharge current $I_{d2}$ in cage II for different environments. (b) Total electron density of electrons for constant discharge current of $I_{d1}=0.5~A$ in Cage I and with a variable discharge current $I_{d2}$ in cage II for different scenarios.}
\end{figure}
\begin{figure}
\includegraphics[width=8.6cm]{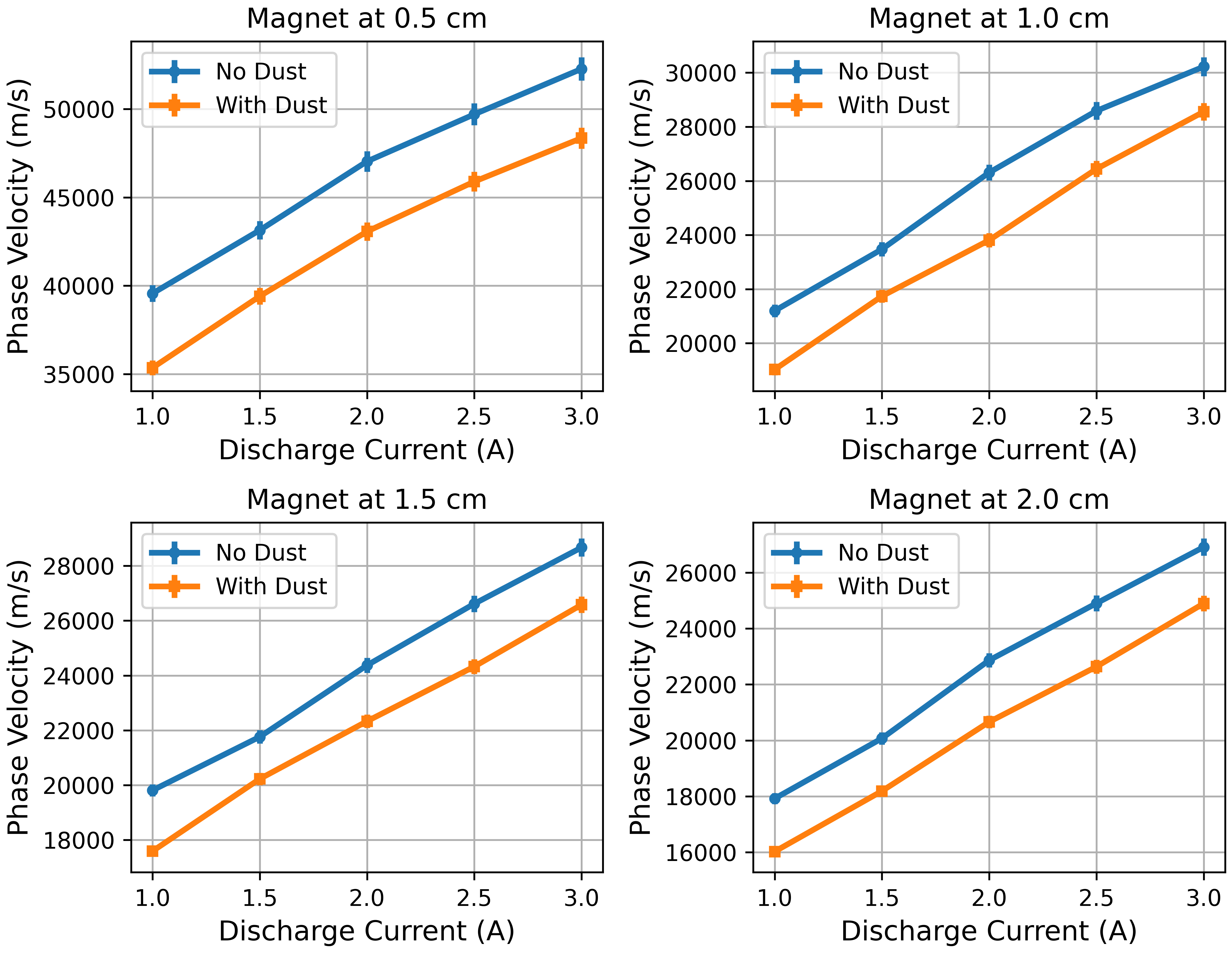}
\centering
\caption{Phase velocity of IAWs at different magnet positions with a constant discharge current of $I_{d1}=0.5~A$ in Cage I and different values of discharge currents in Cage II.}
\end{figure}

The phase velocity of IAWs as a function of the discharge current at various magnet positions is depicted in Fig. 9. It indicates that the phase velocity increases as the discharge current increases, but reduces when dust grains are present. The reason is identical to the one mentioned in the preceding paragraph. It also depicts how the wave phase velocity drop as the magnet is farther away. The strength of the magnetic field at the centre of the plasma column decreases on moving the magnet away from the grid-probe axis. Because of this, the restriction on the movement of the electrons decreases, and the electron density also increases. Consequently, the shielding of the electric field increases, which causes the phase velocity of the wave to decrease.

\subsection{Damping Study of IAWs:}

The damping of the IAWs can be attributed to either wave-particle interaction known as Landau damping or due to the collision of particles in plasma. Landau damping is dominant in a plasma with ion temperature almost equal to the electron temperature \textit{i.e} $T_i \sim T_e$ {\cite{hosseini2011}}. In the present study, $T_e>>T_i$, so the damping of the wave is collisional. The collisional damping of the IAWs mainly arises from i) electron-ion, ii) ion-ion, and iii) ion-neutral collisions {\cite{andersen1968, buti1968}}. In the presence of dust grains, the ion-dust collisions are crucial in damping of the wave {\cite{nakamura1999}}.

\begin{figure}
\includegraphics[width=8.6cm]{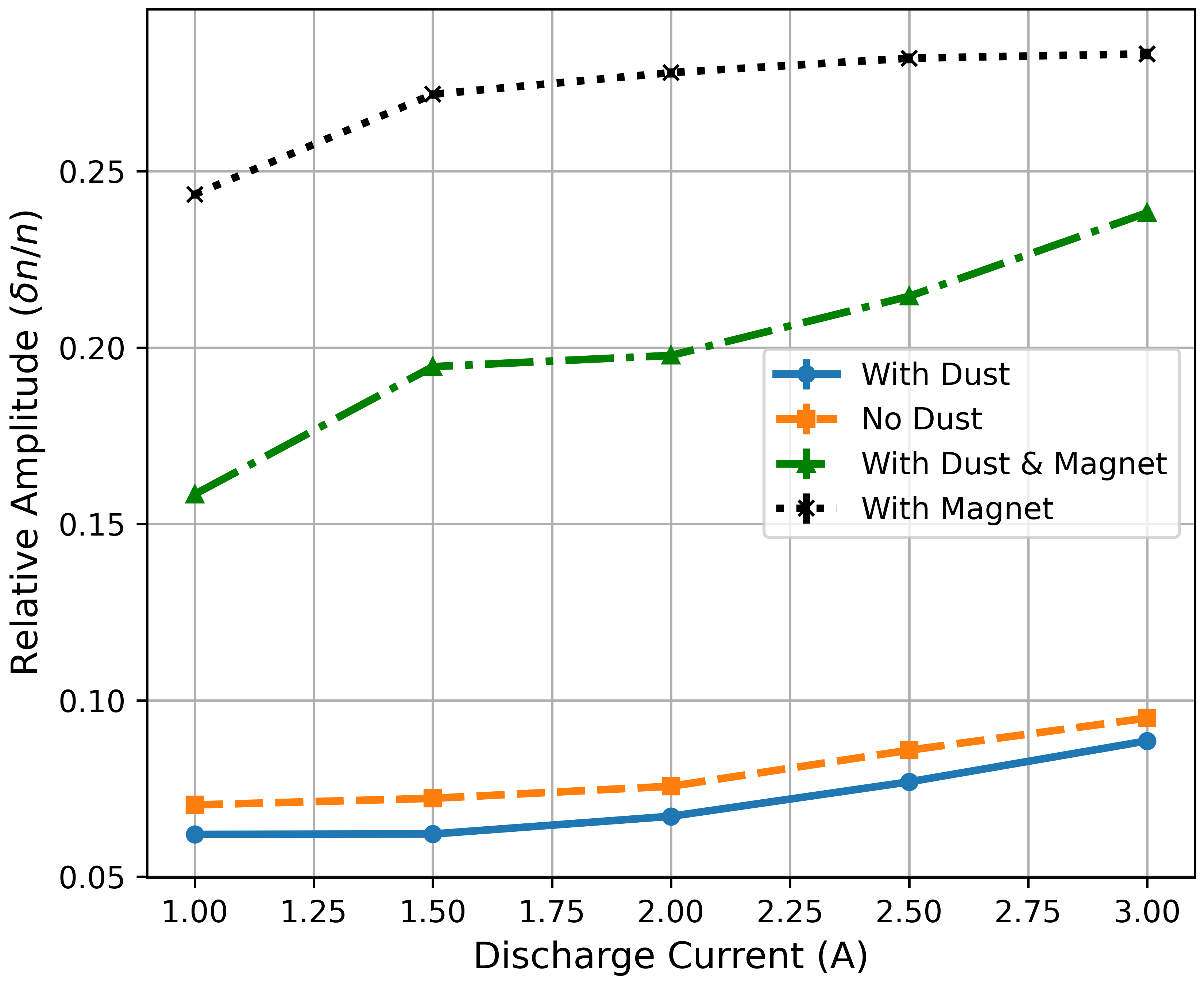}
\centering
\caption{Relative wave amplitude of received signal as a function of discharge currents $I_{d2}$ in Cage II at a constant  discharge current of $I_{d1}=0.5 A$ in Cage I.}
\end{figure}

Fig. 10 shows the relative amplitude of the received signal with discharge current. The amplitude of the received signal is normalized by the amplitude of the applied signal. The figure portrays that the damping of the wave decreases with the increase of discharge current. The plasma discharge current increases due to more and more ionization, which causes the neutral density to decrease at a constant pressure. It reduces the collision frequency of ions and electrons with the neutral background. So, the temperature of the ions and the electrons should increase. From the experimental data, it has been observed that the electron temperature increases with the discharge current. Following the same justification, it may be assumed that ion temperature will also increase slightly. As the electron and ion temperature rises, the electron-ion and ion-ion collisions (coulomb collision) will reduce. Hence,  the collisions responsible for the damping of the IAWs will decrease, and consequently, the damping will be less.

The damping of the wave will further enhance when dust grains are introduced into the plasma. This additional damping can be attributed to ion-dust collisions. On placing the external magnetic field in the absence of the dust grains, the damping of the wave reduces significantly.  The curvature of the charged particle can not affect the electron-ion collisions if the Larmor radius of electron ($r_{Le}$) is approximately equal to Debye length (${\lambda}_D$), and electron cyclotron frequency ($\Omega_e$) is of the order of electron plasma frequency ($\omega_e$) {\cite{geller1997}}. In the present scenario, $r_{Le}\approx{\lambda}_D$ and ${\Omega}_e\approx{\omega}_e$, which suggest that the magnetic field strength used in this study can not affect the electron-ion collisions. The gyro-radius of the ions ($r_{Li}$) is greater than the electron gyro-radius ($r_{Le}$), consequently the curvature of the ion's trajectory is smaller than the electron's trajectory, which should not affect the ion-ion collisions too. On the contrary, the presence of the magnetic field will reduce the ion-neutral collision frequency {\cite{imazu1981}}, which in turn reduces the damping of IAWs. The dampening of the wave rises when  dust particles are introduced while the magnetic field is still present in the system.

Fig. 11 shows the relative wave amplitudes of the received signal for various values of discharge currents at different magnet positions. It appears that the damping decreases with the increase of discharge current, and with the introduction of dust particles, the damping enhances. It also depicts that the damping increases on moving the magnet away from the grid-probe axis. Imazu {\cite{imazu1981}} has shown that for magnetic field strength, in the range of 1.5 kG to 0.1 kG, the ion-neutral collision frequency increases almost linearly. The effective field strength for this study varies from 1.43 kG to 0.24 kG, which lies within the range discussed above, consequently the ion-neutral collision increases followed by the damping.

\begin{figure}
\includegraphics[width=8.6cm]{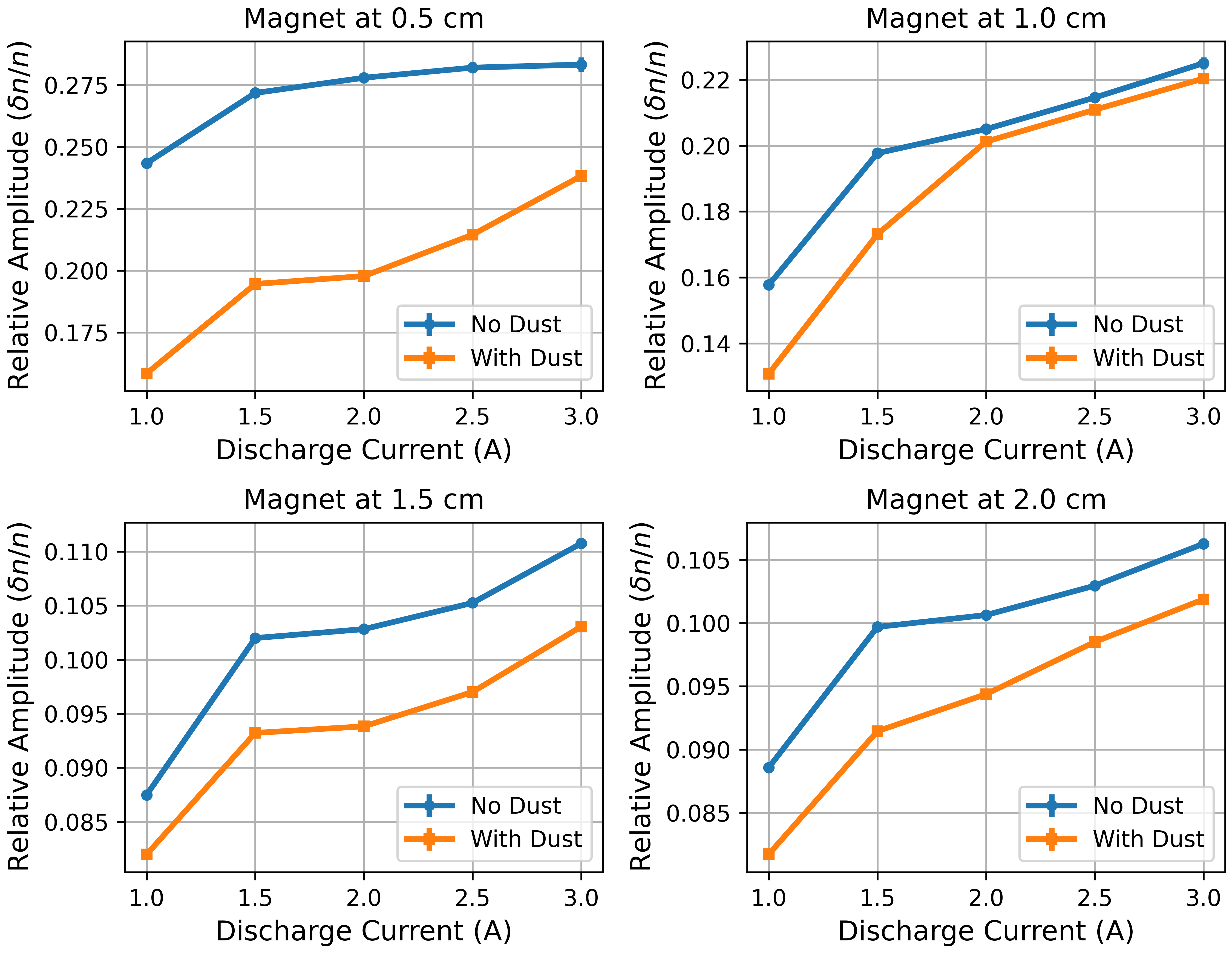}
\centering
\caption{Relative amplitude of wave at different magnet positions as a function of discharge currents $I_{d2}$ in Cage II keeping the discharge currents in Cage I at a constant value of $I_{d1}=0.5~A$.}
\end{figure}

\section{Conclusions}
The effect of the external magnetic fields on the characteristics of the IAWs in a two-temperature hydrogen plasma has been investigated. The magnetic field has a considerable impact on the propagation and dampening of IAWs. 

1. The presence of a magnetic field in a direction perpendicular to the propagation of IAWs enhances the phase velocity and reduces the damping of the IAWs. The phase velocity increases due to the restriction on the movement of electrons and the reduction of electron density. The electron density decreases because of the diffusion along the field direction. As a result, the electric field generated by the ion bunches is less shielded. The damping is weakened owing to the decrease of ion-neutral collisions in the presence of the magnetic field. 

2. The wave phase velocity is lower in the presence of the magnetic field  and dust particles than it is when the magnetic field is present alone, which is due to secondary emissions and scattering of the electrons from the dust grains.

3. When the magnet is gradually moved away from the grid probe axis, the damping of the waves increases, but the phase velocity decreases. The damping grows as the ion-neutral collisions increase with the lowering of the field strength. On the contrary, the restrictions on the movement of the electrons decrease, and electron density also increases with the reduction of the field strength. Due to this, the shielding of the electric field associated with the ion bunches increases, and consequently, the phase velocity decreases.

4. When dust particles are introduced in the plasma in the absence of magnetic field, the phase velocity of the IAWs decreases. The secondary electron emission from the tungsten (W) dust, caused by the energetic group of electrons, in the range of $10-12~eV$ is responsible for the reduction in velocity. This phenomenon is possible only if there are energetic groups of electrons in the plasma.  

5. The damping of the wave in the presence of dust grains increases for all the plasma environments explored in the present study. This damping is directly associated with ion-dust collisions.

In the presence of the external magnetic field, the ion-neutral collisions seem to be responsible for the change in damping of the wave. Therefore, from the damping of the IAWs, it may be possible to determine the ion-neutral collision frequency. The study can be used to confirm the secondary emissions of electrons from dust grains in a two-electron temperature plasma. The results of this study will be helpful to use IAWs as a diagnostic tool for probing plasma parameters in the presence of magnetic field. It is worth mentioning that the results obtained are independent of the present experimental set-up. Such observations can be replicated in any experimental system containing two electron groups with different energies. The next step of the study will be to create a theoretical model to determine the dispersion relation of the IAWs and eventually describe the current findings.

%*****************************************************************************************    
\section*{DATA AVAILABILITY STATEMENT}
The data that support the findings of this study are available from the corresponding author upon reasonable request.

\section*{Acknowledgement}
The authors would like to thank Mr. Gobinda Dev Sarma, and Mr. Shomsoz Zaman for their technical helps during the experiment.

\nocite{*}

\bibliography{apssamp}% Produces the bibliography via BibTeX.

\providecommand{\noopsort}[1]{}\providecommand{\singleletter}[1]{#1}%
\begin{thebibliography}{45}%
\makeatletter
\providecommand \@ifxundefined [1]{%
 \@ifx{#1\undefined}
}%
\providecommand \@ifnum [1]{%
 \ifnum #1\expandafter \@firstoftwo
 \else \expandafter \@secondoftwo
 \fi
}%
\providecommand \@ifx [1]{%
 \ifx #1\expandafter \@firstoftwo
 \else \expandafter \@secondoftwo
 \fi
}%
\providecommand \natexlab [1]{#1}%
\providecommand \enquote  [1]{``#1''}%
\providecommand \bibnamefont  [1]{#1}%
\providecommand \bibfnamefont [1]{#1}%
\providecommand \citenamefont [1]{#1}%
\providecommand \href@noop [0]{\@secondoftwo}%
\providecommand \href [0]{\begingroup \@sanitize@url \@href}%
\providecommand \@href[1]{\@@startlink{#1}\@@href}%
\providecommand \@@href[1]{\endgroup#1\@@endlink}%
\providecommand \@sanitize@url [0]{\catcode `\\12\catcode `\$12\catcode
  `\&12\catcode `\#12\catcode `\^12\catcode `\_12\catcode `\%12\relax}%
\providecommand \@@startlink[1]{}%
\providecommand \@@endlink[0]{}%
\providecommand \url  [0]{\begingroup\@sanitize@url \@url }%
\providecommand \@url [1]{\endgroup\@href {#1}{\urlprefix }}%
\providecommand \urlprefix  [0]{URL }%
\providecommand \Eprint [0]{\href }%
\providecommand \doibase [0]{https://doi.org/}%
\providecommand \selectlanguage [0]{\@gobble}%
\providecommand \bibinfo  [0]{\@secondoftwo}%
\providecommand \bibfield  [0]{\@secondoftwo}%
\providecommand \translation [1]{[#1]}%
\providecommand \BibitemOpen [0]{}%
\providecommand \bibitemStop [0]{}%
\providecommand \bibitemNoStop [0]{.\EOS\space}%
\providecommand \EOS [0]{\spacefactor3000\relax}%
\providecommand \BibitemShut  [1]{\csname bibitem#1\endcsname}%
\let\auto@bib@innerbib\@empty
%</preamble>
\bibitem [{\citenamefont {Hosseini~Jenab}\ \emph {et~al.}(2011)\citenamefont
  {Hosseini~Jenab}, \citenamefont {Kourakis},\ and\ \citenamefont
  {Abbasi}}]{hosseini2011}%
  \BibitemOpen
  \bibfield  {author} {\bibinfo {author} {\bibfnamefont {S.}~\bibnamefont
  {Hosseini~Jenab}}, \bibinfo {author} {\bibfnamefont {I.}~\bibnamefont
  {Kourakis}},\ and\ \bibinfo {author} {\bibfnamefont {H.}~\bibnamefont
  {Abbasi}},\ }\bibfield  {title} {\bibinfo {title} {Fully kinetic simulation
  of ion acoustic and dust-ion acoustic waves},\ }\href@noop {} {\bibfield
  {journal} {\bibinfo  {journal} {Physics of Plasmas}\ }\textbf {\bibinfo
  {volume} {18}},\ \bibinfo {pages} {073703} (\bibinfo {year}
  {2011})}\BibitemShut {NoStop}%
\bibitem [{\citenamefont {Langmuir}(1929)}]{tonks1929}%
  \BibitemOpen
  \bibfield  {author} {\bibinfo {author} {\bibfnamefont {I.}~\bibnamefont
  {Langmuir}},\ }\bibfield  {title} {\bibinfo {title} {Tonks., l. ion
  oscillations in a warm plasma},\ }\href@noop {} {\bibfield  {journal}
  {\bibinfo  {journal} {Phys. Rev}\ }\textbf {\bibinfo {volume} {33}},\
  \bibinfo {pages} {195} (\bibinfo {year} {1929})}\BibitemShut {NoStop}%
\bibitem [{\citenamefont {Revans}(1933)}]{Revans1933}%
  \BibitemOpen
  \bibfield  {author} {\bibinfo {author} {\bibfnamefont {R.}~\bibnamefont
  {Revans}},\ }\bibfield  {title} {\bibinfo {title} {The transmission of waves
  through an ionized gas},\ }\href@noop {} {\bibfield  {journal} {\bibinfo
  {journal} {Physical Review}\ }\textbf {\bibinfo {volume} {44}},\ \bibinfo
  {pages} {798} (\bibinfo {year} {1933})}\BibitemShut {NoStop}%
\bibitem [{\citenamefont {Mahmood}\ \emph {et~al.}(2003)\citenamefont
  {Mahmood}, \citenamefont {Mushtaq},\ and\ \citenamefont
  {Saleem}}]{mahmood2003}%
  \BibitemOpen
  \bibfield  {author} {\bibinfo {author} {\bibfnamefont {S.}~\bibnamefont
  {Mahmood}}, \bibinfo {author} {\bibfnamefont {A.}~\bibnamefont {Mushtaq}},\
  and\ \bibinfo {author} {\bibfnamefont {H.}~\bibnamefont {Saleem}},\
  }\bibfield  {title} {\bibinfo {title} {Ion acoustic solitary wave in
  homogeneous magnetized electron-positron-ion plasmas},\ }\href@noop {}
  {\bibfield  {journal} {\bibinfo  {journal} {New Journal of Physics}\ }\textbf
  {\bibinfo {volume} {5}},\ \bibinfo {pages} {28} (\bibinfo {year}
  {2003})}\BibitemShut {NoStop}%
\bibitem [{\citenamefont {Thomson}\ and\ \citenamefont
  {Thomson}(1933)}]{thomson}%
  \BibitemOpen
  \bibfield  {author} {\bibinfo {author} {\bibfnamefont {J.}~\bibnamefont
  {Thomson}}\ and\ \bibinfo {author} {\bibfnamefont {G.}~\bibnamefont
  {Thomson}},\ }\href@noop {} {\emph {\bibinfo {title} {Conduction of
  Electricity Through Gases}}},\ \bibinfo {number} {v. 2}\ (\bibinfo
  {publisher} {Cambridge University Press},\ \bibinfo {year}
  {1933})\BibitemShut {NoStop}%
\bibitem [{\citenamefont {Wong}\ \emph {et~al.}(1964)\citenamefont {Wong},
  \citenamefont {Motley},\ and\ \citenamefont {D'Angelo}}]{wong1964}%
  \BibitemOpen
  \bibfield  {author} {\bibinfo {author} {\bibfnamefont {A.~Y.}\ \bibnamefont
  {Wong}}, \bibinfo {author} {\bibfnamefont {R.~W.}\ \bibnamefont {Motley}},\
  and\ \bibinfo {author} {\bibfnamefont {N.}~\bibnamefont {D'Angelo}},\
  }\bibfield  {title} {\bibinfo {title} {Landau damping of ion acoustic waves
  in highly ionized plasmas},\ }\href@noop {} {\bibfield  {journal} {\bibinfo
  {journal} {Phys. Rev.}\ }\textbf {\bibinfo {volume} {133}},\ \bibinfo {pages}
  {A436} (\bibinfo {year} {1964})}\BibitemShut {NoStop}%
\bibitem [{\citenamefont {D'Angelo}\ \emph {et~al.}(1966)\citenamefont
  {D'Angelo}, \citenamefont {Goeler},\ and\ \citenamefont {Ohe}}]{Angelo1966}%
  \BibitemOpen
  \bibfield  {author} {\bibinfo {author} {\bibfnamefont {N.}~\bibnamefont
  {D'Angelo}}, \bibinfo {author} {\bibfnamefont {S.~V.}\ \bibnamefont
  {Goeler}},\ and\ \bibinfo {author} {\bibfnamefont {T.}~\bibnamefont {Ohe}},\
  }\bibfield  {title} {\bibinfo {title} {Propagation and damping of ion waves
  in a plasma with negative ions},\ }\href@noop {} {\bibfield  {journal}
  {\bibinfo  {journal} {Physics of Fluids}\ }\textbf {\bibinfo {volume} {9}},\
  \bibinfo {pages} {1605} (\bibinfo {year} {1966})}\BibitemShut {NoStop}%
\bibitem [{\citenamefont {Joyce}\ \emph {et~al.}(1969)\citenamefont {Joyce},
  \citenamefont {Lonngren}, \citenamefont {Alexeff},\ and\ \citenamefont
  {Jones}}]{joyce1969}%
  \BibitemOpen
  \bibfield  {author} {\bibinfo {author} {\bibfnamefont {G.}~\bibnamefont
  {Joyce}}, \bibinfo {author} {\bibfnamefont {K.}~\bibnamefont {Lonngren}},
  \bibinfo {author} {\bibfnamefont {I.}~\bibnamefont {Alexeff}},\ and\ \bibinfo
  {author} {\bibfnamefont {W.}~\bibnamefont {Jones}},\ }\bibfield  {title}
  {\bibinfo {title} {Dispersion of ion-acoustic waves},\ }\href@noop {}
  {\bibfield  {journal} {\bibinfo  {journal} {The Physics of Fluids}\ }\textbf
  {\bibinfo {volume} {12}},\ \bibinfo {pages} {2592} (\bibinfo {year}
  {1969})}\BibitemShut {NoStop}%
\bibitem [{\citenamefont {Alexeff}\ and\ \citenamefont
  {Jones}(1965)}]{alexeff1965}%
  \BibitemOpen
  \bibfield  {author} {\bibinfo {author} {\bibfnamefont {I.}~\bibnamefont
  {Alexeff}}\ and\ \bibinfo {author} {\bibfnamefont {W.}~\bibnamefont
  {Jones}},\ }\bibfield  {title} {\bibinfo {title} {Collisionless ion-wave
  propagation and the determination of the compression coefficient of plasma
  electrons},\ }\href@noop {} {\bibfield  {journal} {\bibinfo  {journal}
  {Physical Review Letters}\ }\textbf {\bibinfo {volume} {15}},\ \bibinfo
  {pages} {286} (\bibinfo {year} {1965})}\BibitemShut {NoStop}%
\bibitem [{\citenamefont {Tanaca}\ \emph {et~al.}(1966)\citenamefont {Tanaca},
  \citenamefont {Koganei},\ and\ \citenamefont {Hirose}}]{tanaca1966}%
  \BibitemOpen
  \bibfield  {author} {\bibinfo {author} {\bibfnamefont {H.}~\bibnamefont
  {Tanaca}}, \bibinfo {author} {\bibfnamefont {M.}~\bibnamefont {Koganei}},\
  and\ \bibinfo {author} {\bibfnamefont {A.}~\bibnamefont {Hirose}},\
  }\bibfield  {title} {\bibinfo {title} {Dispersion relation of ion waves in
  mercury-vapor discharges},\ }\href@noop {} {\bibfield  {journal} {\bibinfo
  {journal} {Physical Review Letters}\ }\textbf {\bibinfo {volume} {16}},\
  \bibinfo {pages} {1079} (\bibinfo {year} {1966})}\BibitemShut {NoStop}%
\bibitem [{\citenamefont {Popel}\ \emph {et~al.}(1995)\citenamefont {Popel},
  \citenamefont {Vladimirov},\ and\ \citenamefont {Shukla}}]{popel1995}%
  \BibitemOpen
  \bibfield  {author} {\bibinfo {author} {\bibfnamefont {S.}~\bibnamefont
  {Popel}}, \bibinfo {author} {\bibfnamefont {S.}~\bibnamefont {Vladimirov}},\
  and\ \bibinfo {author} {\bibfnamefont {P.}~\bibnamefont {Shukla}},\
  }\bibfield  {title} {\bibinfo {title} {Ion-acoustic solitons in
  electron--positron--ion plasmas},\ }\href@noop {} {\bibfield  {journal}
  {\bibinfo  {journal} {Physics of Plasmas}\ }\textbf {\bibinfo {volume} {2}},\
  \bibinfo {pages} {716} (\bibinfo {year} {1995})}\BibitemShut {NoStop}%
\bibitem [{\citenamefont {Taylor}\ \emph {et~al.}(1970)\citenamefont {Taylor},
  \citenamefont {Baker},\ and\ \citenamefont {Ikezi}}]{taylor1970}%
  \BibitemOpen
  \bibfield  {author} {\bibinfo {author} {\bibfnamefont {R.}~\bibnamefont
  {Taylor}}, \bibinfo {author} {\bibfnamefont {D.}~\bibnamefont {Baker}},\ and\
  \bibinfo {author} {\bibfnamefont {H.}~\bibnamefont {Ikezi}},\ }\bibfield
  {title} {\bibinfo {title} {Observation of collisionless electrostatic
  shocks},\ }\href@noop {} {\bibfield  {journal} {\bibinfo  {journal} {Physical
  Review Letters}\ }\textbf {\bibinfo {volume} {24}},\ \bibinfo {pages} {206}
  (\bibinfo {year} {1970})}\BibitemShut {NoStop}%
\bibitem [{\citenamefont {Sharma}\ \emph {et~al.}(2020)\citenamefont {Sharma},
  \citenamefont {Adhikari}, \citenamefont {Moulick}, \citenamefont {Kausik},\
  and\ \citenamefont {Saikia}}]{sharma2020}%
  \BibitemOpen
  \bibfield  {author} {\bibinfo {author} {\bibfnamefont {G.}~\bibnamefont
  {Sharma}}, \bibinfo {author} {\bibfnamefont {S.}~\bibnamefont {Adhikari}},
  \bibinfo {author} {\bibfnamefont {R.}~\bibnamefont {Moulick}}, \bibinfo
  {author} {\bibfnamefont {S.}~\bibnamefont {Kausik}},\ and\ \bibinfo {author}
  {\bibfnamefont {B.}~\bibnamefont {Saikia}},\ }\bibfield  {title} {\bibinfo
  {title} {Effect of two temperature electrons in a collisional magnetized
  plasma sheath},\ }\href@noop {} {\bibfield  {journal} {\bibinfo  {journal}
  {Physica Scripta}\ }\textbf {\bibinfo {volume} {95}},\ \bibinfo {pages}
  {035605} (\bibinfo {year} {2020})}\BibitemShut {NoStop}%
\bibitem [{\citenamefont {Shukla}\ and\ \citenamefont
  {Tagare}(1976)}]{shukla1976}%
  \BibitemOpen
  \bibfield  {author} {\bibinfo {author} {\bibfnamefont {P.}~\bibnamefont
  {Shukla}}\ and\ \bibinfo {author} {\bibfnamefont {S.}~\bibnamefont
  {Tagare}},\ }\bibfield  {title} {\bibinfo {title} {Ion-acoustic shock waves
  in multi-electron temperature collisional plasma},\ }\href@noop {} {\bibfield
   {journal} {\bibinfo  {journal} {Physics Letters A}\ }\textbf {\bibinfo
  {volume} {59}},\ \bibinfo {pages} {38} (\bibinfo {year} {1976})}\BibitemShut
  {NoStop}%
\bibitem [{\citenamefont {Jones}\ \emph {et~al.}(1975)\citenamefont {Jones},
  \citenamefont {Lee}, \citenamefont {Gleman},\ and\ \citenamefont
  {Doucet}}]{Jones1975}%
  \BibitemOpen
  \bibfield  {author} {\bibinfo {author} {\bibfnamefont {W.}~\bibnamefont
  {Jones}}, \bibinfo {author} {\bibfnamefont {A.}~\bibnamefont {Lee}}, \bibinfo
  {author} {\bibfnamefont {S.}~\bibnamefont {Gleman}},\ and\ \bibinfo {author}
  {\bibfnamefont {H.}~\bibnamefont {Doucet}},\ }\bibfield  {title} {\bibinfo
  {title} {Propagation of ion-acoustic waves in a two-electron-temperature
  plasma},\ }\href@noop {} {\bibfield  {journal} {\bibinfo  {journal} {Physical
  Review Letters}\ }\textbf {\bibinfo {volume} {35}},\ \bibinfo {pages} {1349}
  (\bibinfo {year} {1975})}\BibitemShut {NoStop}%
\bibitem [{\citenamefont {Merlino}\ \emph {et~al.}(1997)\citenamefont
  {Merlino}, \citenamefont {Barkan}, \citenamefont {Thompson},\ and\
  \citenamefont {D’Angelo}}]{merlino1997}%
  \BibitemOpen
  \bibfield  {author} {\bibinfo {author} {\bibfnamefont {R.}~\bibnamefont
  {Merlino}}, \bibinfo {author} {\bibfnamefont {A.}~\bibnamefont {Barkan}},
  \bibinfo {author} {\bibfnamefont {C.}~\bibnamefont {Thompson}},\ and\
  \bibinfo {author} {\bibfnamefont {N.}~\bibnamefont {D’Angelo}},\ }\bibfield
   {title} {\bibinfo {title} {Plasma phys. control. fusion 39},\ }\href@noop {}
  {\bibfield  {journal} {\bibinfo  {journal} {A421}\ } (\bibinfo {year}
  {1997})}\BibitemShut {NoStop}%
\bibitem [{\citenamefont {Barkan}\ \emph {et~al.}(1996)\citenamefont {Barkan},
  \citenamefont {D'angelo},\ and\ \citenamefont {Merlino}}]{barkan1996}%
  \BibitemOpen
  \bibfield  {author} {\bibinfo {author} {\bibfnamefont {A.}~\bibnamefont
  {Barkan}}, \bibinfo {author} {\bibfnamefont {N.}~\bibnamefont {D'angelo}},\
  and\ \bibinfo {author} {\bibfnamefont {R.}~\bibnamefont {Merlino}},\
  }\bibfield  {title} {\bibinfo {title} {Experiments on ion-acoustic waves in
  dusty plasmas},\ }\href@noop {} {\bibfield  {journal} {\bibinfo  {journal}
  {Planetary and Space Science}\ }\textbf {\bibinfo {volume} {44}},\ \bibinfo
  {pages} {239} (\bibinfo {year} {1996})}\BibitemShut {NoStop}%
\bibitem [{\citenamefont {Nakamura}\ \emph {et~al.}(1999)\citenamefont
  {Nakamura}, \citenamefont {Bailung},\ and\ \citenamefont
  {Shukla}}]{nakamuraPRL}%
  \BibitemOpen
  \bibfield  {author} {\bibinfo {author} {\bibfnamefont {Y.}~\bibnamefont
  {Nakamura}}, \bibinfo {author} {\bibfnamefont {H.}~\bibnamefont {Bailung}},\
  and\ \bibinfo {author} {\bibfnamefont {P.}~\bibnamefont {Shukla}},\
  }\bibfield  {title} {\bibinfo {title} {Observation of ion-acoustic shocks in
  a dusty plasma},\ }\href@noop {} {\bibfield  {journal} {\bibinfo  {journal}
  {Physical review letters}\ }\textbf {\bibinfo {volume} {83}},\ \bibinfo
  {pages} {1602} (\bibinfo {year} {1999})}\BibitemShut {NoStop}%
\bibitem [{\citenamefont {Hadi}\ and\ \citenamefont {Qamar}(2019)}]{hadi2019}%
  \BibitemOpen
  \bibfield  {author} {\bibinfo {author} {\bibfnamefont {F.}~\bibnamefont
  {Hadi}}\ and\ \bibinfo {author} {\bibfnamefont {A.}~\bibnamefont {Qamar}},\
  }\bibfield  {title} {\bibinfo {title} {Kinetic study of dust ion acoustic
  waves in a nonthermal plasma},\ }\href@noop {} {\bibfield  {journal}
  {\bibinfo  {journal} {Journal of the Physical Society of Japan}\ }\textbf
  {\bibinfo {volume} {88}},\ \bibinfo {pages} {034501} (\bibinfo {year}
  {2019})}\BibitemShut {NoStop}%
\bibitem [{\citenamefont {Nakamura}\ and\ \citenamefont
  {Sarma}(2001)}]{nakamura2001}%
  \BibitemOpen
  \bibfield  {author} {\bibinfo {author} {\bibfnamefont {Y.}~\bibnamefont
  {Nakamura}}\ and\ \bibinfo {author} {\bibfnamefont {A.}~\bibnamefont
  {Sarma}},\ }\bibfield  {title} {\bibinfo {title} {Observation of ion-acoustic
  solitary waves in a dusty plasma},\ }\href@noop {} {\bibfield  {journal}
  {\bibinfo  {journal} {Physics of Plasmas}\ }\textbf {\bibinfo {volume} {8}},\
  \bibinfo {pages} {3921} (\bibinfo {year} {2001})}\BibitemShut {NoStop}%
\bibitem [{\citenamefont {Maitra}\ and\ \citenamefont
  {Roychoudhury}(2006)}]{maitra2006}%
  \BibitemOpen
  \bibfield  {author} {\bibinfo {author} {\bibfnamefont {S.}~\bibnamefont
  {Maitra}}\ and\ \bibinfo {author} {\bibfnamefont {R.}~\bibnamefont
  {Roychoudhury}},\ }\bibfield  {title} {\bibinfo {title} {Obliquely
  propagating ion acoustic solitary waves in a dusty plasma in the presence of
  an external magnetic field},\ }\href@noop {} {\bibfield  {journal} {\bibinfo
  {journal} {Physics of plasmas}\ }\textbf {\bibinfo {volume} {13}},\ \bibinfo
  {pages} {112302} (\bibinfo {year} {2006})}\BibitemShut {NoStop}%
\bibitem [{\citenamefont {Nakamura}(2002)}]{nakamura2002}%
  \BibitemOpen
  \bibfield  {author} {\bibinfo {author} {\bibfnamefont {Y.}~\bibnamefont
  {Nakamura}},\ }\bibfield  {title} {\bibinfo {title} {Experiments on
  ion-acoustic shock waves in a dusty plasma},\ }\href@noop {} {\bibfield
  {journal} {\bibinfo  {journal} {Physics of Plasmas}\ }\textbf {\bibinfo
  {volume} {9}},\ \bibinfo {pages} {440} (\bibinfo {year} {2002})}\BibitemShut
  {NoStop}%
\bibitem [{\citenamefont {Farooq}\ and\ \citenamefont
  {Ahmad}(2017)}]{farooq2017}%
  \BibitemOpen
  \bibfield  {author} {\bibinfo {author} {\bibfnamefont {M.}~\bibnamefont
  {Farooq}}\ and\ \bibinfo {author} {\bibfnamefont {M.}~\bibnamefont {Ahmad}},\
  }\bibfield  {title} {\bibinfo {title} {Dust ion acoustic waves in four
  component magnetized dusty plasma with effect of slow rotation and
  superthermal electrons},\ }\href@noop {} {\bibfield  {journal} {\bibinfo
  {journal} {Physics of Plasmas}\ }\textbf {\bibinfo {volume} {24}},\ \bibinfo
  {pages} {123707} (\bibinfo {year} {2017})}\BibitemShut {NoStop}%
\bibitem [{\citenamefont {Hershkowitz}\ and\ \citenamefont
  {Kim}(2008)}]{hershkow2008}%
  \BibitemOpen
  \bibfield  {author} {\bibinfo {author} {\bibfnamefont {N.}~\bibnamefont
  {Hershkowitz}}\ and\ \bibinfo {author} {\bibfnamefont {Y.-C.~G.}\
  \bibnamefont {Kim}},\ }\bibfield  {title} {\bibinfo {title} {Probing plasmas
  with ion acoustic waves},\ }\href@noop {} {\bibfield  {journal} {\bibinfo
  {journal} {Plasma Sources Science and Technology}\ }\textbf {\bibinfo
  {volume} {18}},\ \bibinfo {pages} {014018} (\bibinfo {year}
  {2008})}\BibitemShut {NoStop}%
\bibitem [{\citenamefont {Kakati}\ \emph {et~al.}(2017)\citenamefont {Kakati},
  \citenamefont {Kausik}, \citenamefont {Bandyopadhyay}, \citenamefont
  {Saikia},\ and\ \citenamefont {Kaw}}]{kakati2017}%
  \BibitemOpen
  \bibfield  {author} {\bibinfo {author} {\bibfnamefont {B.}~\bibnamefont
  {Kakati}}, \bibinfo {author} {\bibfnamefont {S.}~\bibnamefont {Kausik}},
  \bibinfo {author} {\bibfnamefont {M.}~\bibnamefont {Bandyopadhyay}}, \bibinfo
  {author} {\bibfnamefont {B.}~\bibnamefont {Saikia}},\ and\ \bibinfo {author}
  {\bibfnamefont {P.}~\bibnamefont {Kaw}},\ }\bibfield  {title} {\bibinfo
  {title} {Development of a novel surface assisted volume negative hydrogen ion
  source},\ }\href@noop {} {\bibfield  {journal} {\bibinfo  {journal}
  {Scientific Reports}\ }\textbf {\bibinfo {volume} {7}},\ \bibinfo {pages} {1}
  (\bibinfo {year} {2017})}\BibitemShut {NoStop}%
\bibitem [{\citenamefont {Saikia}\ \emph {et~al.}(2014)\citenamefont {Saikia},
  \citenamefont {Saikia}, \citenamefont {Goswami},\ and\ \citenamefont
  {Phukan}}]{Saikia2014}%
  \BibitemOpen
  \bibfield  {author} {\bibinfo {author} {\bibfnamefont {P.}~\bibnamefont
  {Saikia}}, \bibinfo {author} {\bibfnamefont {B.~K.}\ \bibnamefont {Saikia}},
  \bibinfo {author} {\bibfnamefont {K.~S.}\ \bibnamefont {Goswami}},\ and\
  \bibinfo {author} {\bibfnamefont {A.}~\bibnamefont {Phukan}},\ }\bibfield
  {title} {\bibinfo {title} {Argon--oxygen dc magnetron discharge plasma probed
  with ion acoustic waves},\ }\href@noop {} {\bibfield  {journal} {\bibinfo
  {journal} {Journal of Vacuum Science \& Technology A: Vacuum, Surfaces, and
  Films}\ }\textbf {\bibinfo {volume} {32}},\ \bibinfo {pages} {031303}
  (\bibinfo {year} {2014})}\BibitemShut {NoStop}%
\bibitem [{\citenamefont {Santoso}\ \emph {et~al.}(2015)\citenamefont
  {Santoso}, \citenamefont {Manoharan}, \citenamefont {O'Byrne},\ and\
  \citenamefont {Corr}}]{santoso2015}%
  \BibitemOpen
  \bibfield  {author} {\bibinfo {author} {\bibfnamefont {J.}~\bibnamefont
  {Santoso}}, \bibinfo {author} {\bibfnamefont {R.}~\bibnamefont {Manoharan}},
  \bibinfo {author} {\bibfnamefont {S.}~\bibnamefont {O'Byrne}},\ and\ \bibinfo
  {author} {\bibfnamefont {C.}~\bibnamefont {Corr}},\ }\bibfield  {title}
  {\bibinfo {title} {Negative hydrogen ion production in a helicon plasma
  source},\ }\href@noop {} {\bibfield  {journal} {\bibinfo  {journal} {Physics
  of Plasmas}\ }\textbf {\bibinfo {volume} {22}},\ \bibinfo {pages} {093513}
  (\bibinfo {year} {2015})}\BibitemShut {NoStop}%
\bibitem [{\citenamefont {Paul}\ \emph {et~al.}(2022)\citenamefont {Paul},
  \citenamefont {Sharma}, \citenamefont {Deka}, \citenamefont {Adhikari},
  \citenamefont {Moulick}, \citenamefont {Kausik},\ and\ \citenamefont
  {Saikia}}]{paul2021}%
  \BibitemOpen
  \bibfield  {author} {\bibinfo {author} {\bibfnamefont {R.}~\bibnamefont
  {Paul}}, \bibinfo {author} {\bibfnamefont {G.}~\bibnamefont {Sharma}},
  \bibinfo {author} {\bibfnamefont {K.}~\bibnamefont {Deka}}, \bibinfo {author}
  {\bibfnamefont {S.}~\bibnamefont {Adhikari}}, \bibinfo {author}
  {\bibfnamefont {R.}~\bibnamefont {Moulick}}, \bibinfo {author} {\bibfnamefont
  {S.~S.}\ \bibnamefont {Kausik}},\ and\ \bibinfo {author} {\bibfnamefont
  {B.}~\bibnamefont {Saikia}},\ }\bibfield  {title} {\bibinfo {title}
  {Experimental study of charging of dust grains in the presence of energetic
  electrons},\ }\href@noop {} {\bibfield  {journal} {\bibinfo  {journal}
  {Plasma Physics and Controlled Fusion}\ }\textbf {\bibinfo {volume} {64}},\
  \bibinfo {pages} {035009} (\bibinfo {year} {2022})}\BibitemShut {NoStop}%
\bibitem [{\citenamefont {Sharma}\ \emph {et~al.}(2022)\citenamefont {Sharma},
  \citenamefont {Deka}, \citenamefont {Paul}, \citenamefont {Adhikari},
  \citenamefont {Moulick}, \citenamefont {Kausik},\ and\ \citenamefont
  {Saikia}}]{sharma2021}%
  \BibitemOpen
  \bibfield  {author} {\bibinfo {author} {\bibfnamefont {G.}~\bibnamefont
  {Sharma}}, \bibinfo {author} {\bibfnamefont {K.}~\bibnamefont {Deka}},
  \bibinfo {author} {\bibfnamefont {R.}~\bibnamefont {Paul}}, \bibinfo {author}
  {\bibfnamefont {S.}~\bibnamefont {Adhikari}}, \bibinfo {author}
  {\bibfnamefont {R.}~\bibnamefont {Moulick}}, \bibinfo {author} {\bibfnamefont
  {S.~S.}\ \bibnamefont {Kausik}},\ and\ \bibinfo {author} {\bibfnamefont
  {B.}~\bibnamefont {Saikia}},\ }\bibfield  {title} {\bibinfo {title}
  {Experimental study on controlled production of two-electron temperature
  plasma},\ }\href@noop {} {\bibfield  {journal} {\bibinfo  {journal} {Plasma
  Sources Science and Technology}\ } (\bibinfo {year} {2022})}\BibitemShut
  {NoStop}%
\bibitem [{\citenamefont {Pustylnik}\ \emph {et~al.}(2006)\citenamefont
  {Pustylnik}, \citenamefont {Ohno},\ and\ \citenamefont
  {Takamura}}]{pustylnik2006}%
  \BibitemOpen
  \bibfield  {author} {\bibinfo {author} {\bibfnamefont {M.}~\bibnamefont
  {Pustylnik}}, \bibinfo {author} {\bibfnamefont {N.}~\bibnamefont {Ohno}},\
  and\ \bibinfo {author} {\bibfnamefont {S.}~\bibnamefont {Takamura}},\
  }\bibfield  {title} {\bibinfo {title} {Control of energetic electron
  component in a magnetically confined diffusion ar plasma},\ }\href@noop {}
  {\bibfield  {journal} {\bibinfo  {journal} {Japanese journal of applied
  physics}\ }\textbf {\bibinfo {volume} {45}},\ \bibinfo {pages} {926}
  (\bibinfo {year} {2006})}\BibitemShut {NoStop}%
\bibitem [{\citenamefont {Pilling}\ and\ \citenamefont
  {Carnegie}(2007)}]{pilling2007}%
  \BibitemOpen
  \bibfield  {author} {\bibinfo {author} {\bibfnamefont {L.~S.}\ \bibnamefont
  {Pilling}}\ and\ \bibinfo {author} {\bibfnamefont {D.}~\bibnamefont
  {Carnegie}},\ }\bibfield  {title} {\bibinfo {title} {Validating experimental
  and theoretical langmuir probe analyses},\ }\href@noop {} {\bibfield
  {journal} {\bibinfo  {journal} {Plasma Sources Science and Technology}\
  }\textbf {\bibinfo {volume} {16}},\ \bibinfo {pages} {570} (\bibinfo {year}
  {2007})}\BibitemShut {NoStop}%
\bibitem [{\citenamefont {Rao}\ and\ \citenamefont {Mamun}(2001)}]{mamun2001}%
  \BibitemOpen
  \bibfield  {author} {\bibinfo {author} {\bibfnamefont {N.}~\bibnamefont
  {Rao}}\ and\ \bibinfo {author} {\bibfnamefont {A.}~\bibnamefont {Mamun}},\
  }\href@noop {} {\emph {\bibinfo {title} {Introduction to Dusty Plasma
  Physics}}}\ (\bibinfo  {publisher} {CRC Press},\ \bibinfo {year}
  {2001})\BibitemShut {NoStop}%
\bibitem [{\citenamefont {Pitts}\ and\ \citenamefont
  {Stangeby}(1990)}]{pitts1990}%
  \BibitemOpen
  \bibfield  {author} {\bibinfo {author} {\bibfnamefont {R.}~\bibnamefont
  {Pitts}}\ and\ \bibinfo {author} {\bibfnamefont {P.}~\bibnamefont
  {Stangeby}},\ }\bibfield  {title} {\bibinfo {title} {Experimental tests of
  langmuir probe theory for strong magnetic fields},\ }\href@noop {} {\bibfield
   {journal} {\bibinfo  {journal} {Plasma physics and controlled fusion}\
  }\textbf {\bibinfo {volume} {32}},\ \bibinfo {pages} {1237} (\bibinfo {year}
  {1990})}\BibitemShut {NoStop}%
\bibitem [{\citenamefont {Stangeby}(1982)}]{stangeby1982}%
  \BibitemOpen
  \bibfield  {author} {\bibinfo {author} {\bibfnamefont {P.}~\bibnamefont
  {Stangeby}},\ }\bibfield  {title} {\bibinfo {title} {Effect of bias on
  trapping probes and bolometers for tokamak edge diagnosis},\ }\href@noop {}
  {\bibfield  {journal} {\bibinfo  {journal} {Journal of Physics D: Applied
  Physics}\ }\textbf {\bibinfo {volume} {15}},\ \bibinfo {pages} {1007}
  (\bibinfo {year} {1982})}\BibitemShut {NoStop}%
\bibitem [{\citenamefont {Tagle}\ \emph {et~al.}(1987)\citenamefont {Tagle},
  \citenamefont {Stangeby},\ and\ \citenamefont {Erents}}]{tagle1987}%
  \BibitemOpen
  \bibfield  {author} {\bibinfo {author} {\bibfnamefont {J.}~\bibnamefont
  {Tagle}}, \bibinfo {author} {\bibfnamefont {P.}~\bibnamefont {Stangeby}},\
  and\ \bibinfo {author} {\bibfnamefont {S.}~\bibnamefont {Erents}},\
  }\bibfield  {title} {\bibinfo {title} {Errors in measuring electron
  temperatures using a single langmuir probe in a magnetic field},\ }\href@noop
  {} {\bibfield  {journal} {\bibinfo  {journal} {Plasma Physics and Controlled
  Fusion}\ }\textbf {\bibinfo {volume} {29}},\ \bibinfo {pages} {297} (\bibinfo
  {year} {1987})}\BibitemShut {NoStop}%
\bibitem [{\citenamefont {Stanojevi{\'c}}\ \emph {et~al.}(1994)\citenamefont
  {Stanojevi{\'c}}, \citenamefont {{\v{C}}er{\v{c}}ek}, \citenamefont
  {Gyergyek},\ and\ \citenamefont {Jeli{\'c}}}]{stanojevic1994}%
  \BibitemOpen
  \bibfield  {author} {\bibinfo {author} {\bibfnamefont {M.}~\bibnamefont
  {Stanojevi{\'c}}}, \bibinfo {author} {\bibfnamefont {M.}~\bibnamefont
  {{\v{C}}er{\v{c}}ek}}, \bibinfo {author} {\bibfnamefont {T.}~\bibnamefont
  {Gyergyek}},\ and\ \bibinfo {author} {\bibfnamefont {N.}~\bibnamefont
  {Jeli{\'c}}},\ }\bibfield  {title} {\bibinfo {title} {Interpretation of a
  planar langmuir probe current—voltage characteristic in a strong magnetic
  field},\ }\href@noop {} {\bibfield  {journal} {\bibinfo  {journal}
  {Contributions to Plasma Physics}\ }\textbf {\bibinfo {volume} {34}},\
  \bibinfo {pages} {607} (\bibinfo {year} {1994})}\BibitemShut {NoStop}%
\bibitem [{\citenamefont {Erents}\ \emph {et~al.}(1986)\citenamefont {Erents},
  \citenamefont {Tagle}, \citenamefont {McCracken}, \citenamefont {Stangeby},\
  and\ \citenamefont {De~Kock}}]{erents1986}%
  \BibitemOpen
  \bibfield  {author} {\bibinfo {author} {\bibfnamefont {S.}~\bibnamefont
  {Erents}}, \bibinfo {author} {\bibfnamefont {J.}~\bibnamefont {Tagle}},
  \bibinfo {author} {\bibfnamefont {G.}~\bibnamefont {McCracken}}, \bibinfo
  {author} {\bibfnamefont {P.}~\bibnamefont {Stangeby}},\ and\ \bibinfo
  {author} {\bibfnamefont {L.}~\bibnamefont {De~Kock}},\ }\bibfield  {title}
  {\bibinfo {title} {Probe measurements of the density and temperature profiles
  in the jet plasma boundary},\ }\href@noop {} {\bibfield  {journal} {\bibinfo
  {journal} {Nuclear fusion}\ }\textbf {\bibinfo {volume} {26}},\ \bibinfo
  {pages} {1591} (\bibinfo {year} {1986})}\BibitemShut {NoStop}%
\bibitem [{\citenamefont {Usoltceva}\ \emph {et~al.}(2018)\citenamefont
  {Usoltceva}, \citenamefont {Faudot}, \citenamefont {Devaux}, \citenamefont
  {Heuraux}, \citenamefont {Ledig}, \citenamefont {Zadvitskiy}, \citenamefont
  {Ochoukov}, \citenamefont {Cromb{\'e}},\ and\ \citenamefont
  {Noterdaeme}}]{maria2018}%
  \BibitemOpen
  \bibfield  {author} {\bibinfo {author} {\bibfnamefont {M.}~\bibnamefont
  {Usoltceva}}, \bibinfo {author} {\bibfnamefont {E.}~\bibnamefont {Faudot}},
  \bibinfo {author} {\bibfnamefont {S.}~\bibnamefont {Devaux}}, \bibinfo
  {author} {\bibfnamefont {S.}~\bibnamefont {Heuraux}}, \bibinfo {author}
  {\bibfnamefont {J.}~\bibnamefont {Ledig}}, \bibinfo {author} {\bibfnamefont
  {G.~V.}\ \bibnamefont {Zadvitskiy}}, \bibinfo {author} {\bibfnamefont
  {R.}~\bibnamefont {Ochoukov}}, \bibinfo {author} {\bibfnamefont
  {K.}~\bibnamefont {Cromb{\'e}}},\ and\ \bibinfo {author} {\bibfnamefont
  {J.-M.}\ \bibnamefont {Noterdaeme}},\ }\bibfield  {title} {\bibinfo {title}
  {Effective collecting area of a cylindrical langmuir probe in magnetized
  plasma},\ }\href@noop {} {\bibfield  {journal} {\bibinfo  {journal} {Physics
  of Plasmas}\ }\textbf {\bibinfo {volume} {25}},\ \bibinfo {pages} {063518}
  (\bibinfo {year} {2018})}\BibitemShut {NoStop}%
\bibitem [{\citenamefont {Merlino}(2007)}]{merlino2007}%
  \BibitemOpen
  \bibfield  {author} {\bibinfo {author} {\bibfnamefont {R.~L.}\ \bibnamefont
  {Merlino}},\ }\bibfield  {title} {\bibinfo {title} {Understanding langmuir
  probe current-voltage characteristics},\ }\href@noop {} {\bibfield  {journal}
  {\bibinfo  {journal} {American Journal of Physics}\ }\textbf {\bibinfo
  {volume} {75}},\ \bibinfo {pages} {1078} (\bibinfo {year}
  {2007})}\BibitemShut {NoStop}%
\bibitem [{\citenamefont {Mamun}\ and\ \citenamefont
  {Shukla}(2003)}]{mamun2003}%
  \BibitemOpen
  \bibfield  {author} {\bibinfo {author} {\bibfnamefont {A.}~\bibnamefont
  {Mamun}}\ and\ \bibinfo {author} {\bibfnamefont {P.}~\bibnamefont {Shukla}},\
  }\bibfield  {title} {\bibinfo {title} {Charging of dust grains in a plasma
  with negative ions},\ }\href@noop {} {\bibfield  {journal} {\bibinfo
  {journal} {Physics of Plasmas}\ }\textbf {\bibinfo {volume} {10}},\ \bibinfo
  {pages} {1518} (\bibinfo {year} {2003})}\BibitemShut {NoStop}%
\bibitem [{\citenamefont {Andersen}\ \emph {et~al.}(1968)\citenamefont
  {Andersen}, \citenamefont {D'Angelo}, \citenamefont {Jensen}, \citenamefont
  {Michelsen},\ and\ \citenamefont {Nielsen}}]{andersen1968}%
  \BibitemOpen
  \bibfield  {author} {\bibinfo {author} {\bibfnamefont {H.}~\bibnamefont
  {Andersen}}, \bibinfo {author} {\bibfnamefont {N.}~\bibnamefont {D'Angelo}},
  \bibinfo {author} {\bibfnamefont {V.~O.}\ \bibnamefont {Jensen}}, \bibinfo
  {author} {\bibfnamefont {P.}~\bibnamefont {Michelsen}},\ and\ \bibinfo
  {author} {\bibfnamefont {P.}~\bibnamefont {Nielsen}},\ }\bibfield  {title}
  {\bibinfo {title} {Effects of ion-atom collisions on the propagation and
  damping of ion-acoustic waves},\ }\href@noop {} {\bibfield  {journal}
  {\bibinfo  {journal} {The Physics of Fluids}\ }\textbf {\bibinfo {volume}
  {11}},\ \bibinfo {pages} {1177} (\bibinfo {year} {1968})}\BibitemShut
  {NoStop}%
\bibitem [{\citenamefont {Buti}(1968)}]{buti1968}%
  \BibitemOpen
  \bibfield  {author} {\bibinfo {author} {\bibfnamefont {B.}~\bibnamefont
  {Buti}},\ }\bibfield  {title} {\bibinfo {title} {Ion acoustic waves in a
  collisional plasma},\ }\href@noop {} {\bibfield  {journal} {\bibinfo
  {journal} {Physical Review}\ }\textbf {\bibinfo {volume} {165}},\ \bibinfo
  {pages} {195} (\bibinfo {year} {1968})}\BibitemShut {NoStop}%
\bibitem [{\citenamefont {Nakamura}\ and\ \citenamefont
  {Bailung}(1999)}]{nakamura1999}%
  \BibitemOpen
  \bibfield  {author} {\bibinfo {author} {\bibfnamefont {Y.}~\bibnamefont
  {Nakamura}}\ and\ \bibinfo {author} {\bibfnamefont {H.}~\bibnamefont
  {Bailung}},\ }\bibfield  {title} {\bibinfo {title} {A dusty double plasma
  device},\ }\href@noop {} {\bibfield  {journal} {\bibinfo  {journal} {Review
  of scientific instruments}\ }\textbf {\bibinfo {volume} {70}},\ \bibinfo
  {pages} {2345} (\bibinfo {year} {1999})}\BibitemShut {NoStop}%
\bibitem [{\citenamefont {Geller}\ and\ \citenamefont
  {Weisheit}(1997)}]{geller1997}%
  \BibitemOpen
  \bibfield  {author} {\bibinfo {author} {\bibfnamefont {D.~K.}\ \bibnamefont
  {Geller}}\ and\ \bibinfo {author} {\bibfnamefont {J.~C.}\ \bibnamefont
  {Weisheit}},\ }\bibfield  {title} {\bibinfo {title} {Classical electron-ion
  scattering in strongly magnetized plasmas. i. a generalized coulomb
  logarithm},\ }\href@noop {} {\bibfield  {journal} {\bibinfo  {journal}
  {Physics of Plasmas}\ }\textbf {\bibinfo {volume} {4}},\ \bibinfo {pages}
  {4258} (\bibinfo {year} {1997})}\BibitemShut {NoStop}%
\bibitem [{\citenamefont {Imazu}(1981)}]{imazu1981}%
  \BibitemOpen
  \bibfield  {author} {\bibinfo {author} {\bibfnamefont {S.}~\bibnamefont
  {Imazu}},\ }\bibfield  {title} {\bibinfo {title} {Collision frequency of
  charged particles in a weekly ionized gas in a strong magnetic field},\
  }\href@noop {} {\bibfield  {journal} {\bibinfo  {journal} {Physical Review
  A}\ }\textbf {\bibinfo {volume} {23}},\ \bibinfo {pages} {2644} (\bibinfo
  {year} {1981})}\BibitemShut {NoStop}%
\end{thebibliography}%

\end{document}